\documentclass[]{interact}

\usepackage{epstopdf}
\usepackage{subfigure}

\usepackage{verbatim}
\usepackage{amsmath}
\usepackage{amsfonts}
\usepackage{amssymb}
\usepackage{multirow}
\usepackage{enumitem}
\usepackage{color}
\usepackage{hyperref}
\usepackage{mathrsfs}
\usepackage{algorithm}
\usepackage{setspace}

\usepackage{natbib}

\theoremstyle{plain}
\newtheorem{theorem}{Theorem}[section]

\newtheorem{assumption}[theorem]{Assumption}

\theoremstyle{definition}

\theoremstyle{remark}

\newcommand\bbe{\mbox{\boldmath${\beta}$}}

\newcommand\beg{\mbox{\boldmath${\varepsilon}$}}

\newcommand\bGa{\mbox{\boldmath${\Gamma}$}}

\newcommand\bPhi{\mbox{\boldmath${\Phi}$}}

\newcommand\bG{{\bf G}}

\newcommand\bM{{\bf M}}
\newcommand\mcM{{\mathcal M}}

\newcommand\mbR{{\mathbb R}}

\newcommand\bV{{\bf V}}

\newcommand\bW{{\bf W}}

\newcommand\bY{{\bf Y}}

\newcommand\bZ{{\bf Z}}

\usepackage{setspace}

\begin{document}

\title{Manifold Data Analysis with Applications to High-Frequency 3D Imaging}

\author{
\name{Hyun Bin Kang\textsuperscript{a},Matthew Reimherr\textsuperscript{a}, Mark D. Shriver\textsuperscript{b}, and Peter Claes\textsuperscript{c} \thanks{CONTACT Hyun Bin Kang. Email: \href{mailto:huk163@psu.edu}{huk163@psu.edu}; Matthew Reimherr. Email: \href{mailto:mreimherr@psu.edu}{mreimherr@psu.edu}; Mark D. Shriver. Email: \href{mailto:mds17@psu.edu}{mds17@psu.edu}; Peter Claes. Email: \href{mailto:peter.claes@kuleuven.be}{peter.claes@kuleuven.be}}}
\affil{\textsuperscript{a} Department of Statistics, The Pennsylvania State University, University Park, PA, USA;\\ \textsuperscript{b} Department of Anthropology, The Pennsylvania State University, University Park, PA, USA;\\ \textsuperscript{c} Department of Electrical Engineering, KU Leuven, Leuven, Belgium}
}

\maketitle

\begin{abstract}
Many scientific areas are faced with the challenge of extracting information from large, complex, and highly structured data sets. A great deal of modern statistical work focuses on developing tools for handling such data. This paper presents a new subfield of functional data analysis, FDA, which we call {\it Manifold Data Analysis}, or MDA. MDA is concerned with the statistical analysis of samples where one or more variables measured on each unit is a manifold, thus resulting in as many manifolds as we have units. We propose a framework that converts manifolds into functional objects, an efficient 2-step functional principal component method, and a manifold-on-scalar regression model.  This work is motivated by an anthropological application involving 3D facial imaging data, which is discussed extensively throughout the paper.  The proposed framework is used to understand how individual characteristics, such as age and genetic ancestry, influence the shape of the human face.
\end{abstract}

\begin{keywords}
Functional data analysis; Manifold data analysis; Shape analysis; Manifold learning; Functional principal component analysis; Functional regression
\end{keywords}

\doublespacing
\section{Introduction}
Functional data analysis, FDA, has seen a precipitous growth in recent years, due in part to the numerous complex data that have have emerged.  FDA methods exploit what \cite{ramsay:silverman:2005} termed replication and regularization.  In particular, unlike classic nonparametric  smoothing methods, the data usually consist of as many functions as there are statistical units, while the inherent smoothness in the data/parameters can be exploited to achieve greater statistical efficiency than typical multivariate methods (assuming they can even be applied).  The present work is concerned with opening a new avenue for functional data methods; a subbranch of FDA we are calling  \textit{Manifold Data Analysis}, MDA.  In particular, we present an inferential framework when one of the variables being considered is a manifold, and thus we assume we have as many manifolds as we have units.  Our approach utilizes deformation maps from shape analysis and dimension reduction techniques from manifold learning, which allows us to represent each manifold as a Function, which can then be analyzed using FDA techniques.  Currently, shape analysis methods that go beyond an analysis of landmarks is a very active area of research; our hope is that building a connection with FDA will open up exciting avenues for both shape and functional data analysis, while providing powerful and flexible statistical tools.

\subsection{High-Frequency 3D Facial Imaging}
\label{sec:intro:data}

ADAPT (Anthropology, DNA, and the Appearance and Perception of Traits) is an ongoing study at Pennsylvania State University whose aim is to better understand the architecture of human facial diversity using sophisticated biomaging technologies.  
Investigators of ADAPT collected 3D facial images, alongside genetic information, from admixed populations in the US, Brazil, and Cape Verde \citep{claes:etal:2014,claes:hill:shriver:2014}. 
An anthropometric mask \citep{claes:walters:clement:2012} with 7150 quasi-landmarks was mapped onto the original 3D images, establishing a spatially-dense correspondences between faces where all facial images have the exact same number of points and sustain anatomical homology across individuals. A generalized Procrustes superimposition \citep{rohlf:slice:1990} was used to scale and align them.
The participants also provided basic demographic information including gender, age, height, and weight. A sample of 6564 subjects that have been genotyped is used in the present work. The top three plots in Figure \ref{fig:faces} show three examples of the types of facial images that were collected.

There has been extensive research on 3D facial analysis, but the methodologies have been primarily developed in computer science, electrical engineering, and computer vision for face and facial expression recognition \citep{turk:pentland:1991, ahonen:et:al:2006, jain:li:2011, taigman:et:al:2014, huang:et:al:2014}. There is a more limited literature on 3D facial analysis using a statistical framework, and the few existing methods are primarily concerned with classification or estimation based on facial features, not on understanding the influence of different covariates on the 3D faces themselves. \citet{huang:ding:et:al:2014} introduce a local descriptor multi-modal (2D and 3D) for facial gender and ethnicity classification. \citet{xia:et:al:2013} adopt machine learning techniques to find relationships between gender and facial asymmetry. \citet{xia:et:al:2014} examine age effects using a random forest-based regression, but the regression uses features of local shape deformation between facial curves, captured by \textit{Dense Scalar Fields} based on Riemannian shape analysis \citep{drira:et:al:2012}. \citet{kurtek:drira:2015} provide a  statistical shape analysis framework for 3D faces which allows comparison, deformation, and expression and identity classification, but there has not yet been a corresponding regression method developed that directly takes 3D faces as variables. \citet{porromunos:et:al:2014} attempt to represent faces using splines for face recognition but they do not consider them as manifolds. In contrast, this work proposes a novel functional data approach to analyzing 3D faces, which are viewed as smooth manifolds; by constructing 3D facial functional objects, we can utilize existing functional data analysis tools.  Our goal is to build statistical models that elucidate how different covariates affect patterns seen in different faces.  However, we do not reduce the faces down to a few quantitative traits, instead we exploit inherently smooth structures in the face so that they can be analyzed as a whole.

\subsection{Manifolds in Functional Data Analysis}
\label{sec:intro:fda}

Functional Data Analysis \citep{ramsay:silverman:2005,graves:2009,HKbook,kokoszka:reimherr:2017} concerns the analysis of data consisting of random functions. 
It has been used extensively in a variety of fields including geoscience, health studies, kinesiology, and finance to name only a few.
In recent years there has been an increased interest in exploring how to apply functional data techniques when working with different types of manifolds.
\citet{chen:muller:2012:manifold} consider extending functional data techniques to data that are all lying on a single manifold. \citet{elhamifar:vidal:2011} consider clusters of functional data with each cluster lying on a different manifold. \citet{ellingson:2013} consider mean estimation from functional data all lying on a common manifold. \citet{dimeglo:et:al:2014} also try to find a template function using manifold embedding, considering observed functions as variables with values on a single manifold. 
\citet{lila:et:al:2016} provide a smooth principal component analysis algorithm for functions on a two-dimensional manifold.
\cite{ettinger:et:al:2016} map the the internal carotid artery on a planar domain, which is also a manifold that is homeomorphic to a cylinder.
In previous FDA work, all data is assumed to lie on a single manifold or a small number of manifolds.
In this work, however, we assume that each unit is its own manifold, meaning that we have as many manifolds as we have units. High-dimensional 3D imaging is becoming more common in fields such as biology, kinesiology, engineering, and anthropology. In many of these cases, each image is actually a surface sitting in 3D space, i.e. a manifold. However, from image to image, the manifold changes and one cannot assume that all images lie on the same manifold.

\subsection{Overview}
\label{sec:intro:overview}

The remainder of this paper is organized as follows. {In Section \ref{sec:method:fdobj} we present an algorithm for converting manifolds into functional objects, while Section \ref{sec:method:fpca} presents a computationally efficient 2-step Functional Principal Component Analysis (FPCA) algorithm and Section \ref{sec:method:funreg} introduces a manifold-on-scalar regression model.  In Section \ref{sec:adapt} we apply these methods to the ADAPT data, with Section \ref{sec:adapt:fdobj} creating facial functional objects, Section \ref{sec:adapt:fpca} showing the results of the 2-step FPCA, and Section \ref{sec:adapt:funreg} presenting the regression model of gender, age, height, weight, and genetic ancestry on 3D facial functional objects. We conclude the paper with a discussion in Section \ref{sec:discussion}.

\section{Methodology}
\label{sec:method}
In this section we present our approach for handling random samples of manifolds.  Our primary aim is to lay the foundation for analyzing such data using FDA tools.  To accomplish this, we use tools from shape analysis so that each manifold can be associated with a particular deformation map, while we use tools from Manifold learning and FDA to carry out the described computations.  
In Section \ref{sec:method:fdobj} we introduce a statistical framework to embed a sample of manifolds into a real separable Hilbert space, resulting in a sample of functional objects. We then present a computationally efficient 2-step Functional Principal Component Analysis (FPCA) algorithm in Section \ref{sec:method:fpca}. In Section \ref{sec:method:funreg}, we discuss a manifold-on-scalar regression model and hypothesis testing methods for its coefficient functions.

\subsection{Algorithm}
\label{sec:method:fdobj}

In order to ensure the manifolds are comparable and that our algorithm can be applied, we make the following assumptions.
\begin{assumption} Let $\mathscr{Y}_1, \cdots, \mathscr{Y}_N$ be a random sample of manifolds. We assume that, with probability one,
\begin{enumerate}
\item each $\mathscr{Y}_n$ is a compact d-dimensional manifold that is a subset of $\mathbb{R}^D$ with $d < D$,
\item there exists a nonrandom compact d-dimensional $C^1$ Riemannian manifold $\mathcal{M}_0$ such that each $\mathscr{Y}_n$ is homeomorphic to $\mathcal{M}_0$,
\item there exists an atlas for $\mathcal{M}_0$ with a single coordinate chart $\{(\mathcal{M}_0, \psi)\}$ where for any open set $U \subset \mathcal{M}_0$, $\psi: U \rightarrow \psi(U) \subset \mathbb{R}^d$ and ${\bf M}_0 \triangleq \psi(\mathcal{M}_0)$,
\item to each manifold $\mathscr{Y}_n$, there exists a function ${\bf Y}_n: {\bf M}_0 \rightarrow \mathbb{R}^D$ such that ${\bf Y}_n({\bf M}_0) = \mathscr{Y}_n$, up to possibly a set of measure 0,
\item the functions ${\bf Y}_n$ are elements of $L^2({\bf M}_0)$ with probability one, i.e. $\int_{{\bf M}_0} {\bf Y}_n^\top (m) {\bf Y}_n (m) dm < \infty $.
\end{enumerate}
\end{assumption}

The first assumption states that the sample consists of manifolds that are in the same ambient space, $\mathbb{R}^D$. This can be generalized to other spaces, but we do not pursue that here given the scope of our intended applications. The second guarantees that the manifolds are comparable by assuming that they can all be parametrized by a common manifold, $\mathcal{M}_0$. This manifold is assumed to be $C^1$ and Riemannian so that integration over the manifold is well defined \citep{lee:2003}. The third assumption lets us apply manifold learning methods to ``unfold'' $\mathcal{M}_0$ into the simpler set $\bM_0$.  This is primarily for computational convenience, as $\bM_0$ is a an easier domain to work with.  If the third assumption does not hold, then the manifold, $\mcM_0$, cannot be mapped to a set in $\mbR^d$ without tearing it in some way.  Such an assumption is reasonable for our facial applications, but, for example \cite{ettinger:et:al:2016} utilize FDA methods for data measured on the internal carotid atrtery, which is homeomorphic to a cylinder and thus would violate this assumption.  We discuss this further in Section \ref{sec:discussion}.  The fourth assumption simply allows us to identify the manifolds as functions, which are commonly referred to as deformation maps in shape analysis , while the fifth assumption allows us to view those functions as elements of a Hilbert space.

At the heart of our methodology is the view that each manifold can be identified with a function, and then properties such as smoothness can be defined and exploited by utilizing these functions. The major difference between our setting and traditional FDA is that the domain, ${\bf M}_0$, is not observed and must therefore be constructed. Furthermore, it is assumed that the data are smooth with respect to distance along $\mathcal{M}_0$, not along $\mathbb{R}^D$. The framework to construct ${\bf Y}_n: {\bf M}_0 \rightarrow \mathbb{R}^D$ from $\mathscr{Y}_n$ is summarized below. 

\begin{enumerate}
\item Identify a reference manifold $\mathcal{M}_0$.
\item Embed $\mathcal{M}_0$ into a closed bounded connected region of $\mathbb{R}^d$ to construct ${\bf M}_0$.
\item Align $\mathscr{Y}_n$ to $\mathcal{M}_0$ and thus also to ${\bf M}_0$.
\item Construct basis functions from ${\bf M}_0$ to $\mathbb{R}^D$ and express ${\bf Y}_n$ as a linear combination of these functions.
\end{enumerate}

We now discuss each of the steps above. We assume that the raw data is of the form $\{ \mathrm{y}_{npq}: n = 1, \cdots, N; p = 1, \cdots, P; q = 1, \cdots, D \}$, which consists of $P$ $D$-dimensional points observed on manifolds $\mathscr{Y}_1, \cdots, \mathscr{Y}_N$. We assume that each manifold is ultra-densely sampled, and thus can be completely reconstructed with almost no error, which is a common assumption \textit{Dense Functional Data Analysis} \citep{zhang:wang:2016}.

For the first step, the reference manifold $\mathcal{M}_0$ can be taken from an external source, such as previous literature or a previously constructed library of objects, one of the manifolds in the sample, or an average from the sample.
This choice of $\mathcal{M}_0$ is closely related to second and third step, so it needs to be chosen carefully.

Once $\mathcal{M}_0$ is identified, we use manifold learning techniques on $\mathcal{M}_0$ to find ${\bf M}_0$, the embedding of $\mathcal{M}_0$ in to  $\mathbb{R}^d$. The resulting points will be denoted as $\{ \mathrm{m}_{pq}: p = 1, \cdots, P; q = 1, \cdots, d \}$, $P$. Manifold learning has been a very active area of research, and there are a number of popular methods for carrying out this step, including 
Isomap \citep{tenenbaum:desilva:langford:2000}, Laplacian Eigenmaps \citep{belkin:niyogi:2003}, local linear embedding \citep[LLE]{saul:roweis:2003}, local tangent space alignment \citep[LTSA]{zhang:zha:2004}, and Diffusion Map \citep{nadler:lafon:coifman:kevrekidis:2006}. All of these methods aim to find a low-dimensional representation of the given data, but they utilize different strategies towards achieving it. Isomap finds a lower-dimensional embedding that best preserves the geodesic distance between all points, while Laplacian Eigenmaps tries to preserve local distances. LLE seeks to maintain neighborhood distances. LTSA is algorithmically similar to LLE but tries to learn local neighborhood geometry via tangent spaces and aligns them to find the underlying manifold. 
Diffusion map uses a different perspective by considering a random walk ``diffusing'' through the points, and uses a particular eigendecomposition related to that walk to obtain the low dimensional embedding.  
Spanifold \citep{chenouri:2015:spanifold} sets up a tree on the manifold and tries to maintain pairwise distance relationships within the tree while flattening the manifold.  In section \ref{sec:adapt:fdobj} we will compare the performance of these different approaches on the ADAPT data.

We align all $\mathscr{Y}_1, \cdots, \mathscr{Y}_N$ to the reference manifold $\mathcal{M}_0$. In other words, we find a representation of $\mathscr{Y}_n$ as $\{ \tilde{\mathrm{y}}_{npq}: n = 1, \cdots, N; p = 1, \cdots, P; q = 1, \cdots, D \}$ where $\{ \tilde{\mathrm{y}}_{npq}\}$ are $\{ \mathrm{y}_{npq}\}$ aligned to $\mathcal{M}_0$. We assume that $\mathscr{Y}_n$ is homeomorphic to ${\mathcal M}_0$ and thus to ${\bf M}_0$, which allows us to use ${\bf M}_0$ as a common domain for $\mathscr{Y}_1, \cdots, \mathscr{Y}_N$. For this manifold alignment problem, we rely on shape analysis tools such as Procrustes Analysis \citep{mardia:1979}. 

Once the domain, ${\bf M}_0$, is calculated, the next step is to construct ${\bf Y}_n: {\bf M}_0 \rightarrow \mathbb{R}^D$ from $\mathscr{Y}_n$. As functional data are commonly expressed with basis functions, we fix basis functions ${\bf e}_j: {\bf M}_0 \rightarrow \mathbb{R}^D$ and then we express the manifolds in functional data format as 
$$\mathscr{Y}_n \equiv {\bf Y}_n(m) \approx \sum_{j=1}^J b_{nj} {\bf e}_j(m) \quad \quad m \in {\bf M}_0,$$ 
where $b_{nj} \in \mathbb{R}$. The $\hat{b}_{nj}$ can be found by minimizing 
\begin{align}
\sum_{p=1}^P |\tilde{{\bf \mathrm{y}}}_{np} - {\bf Y}_n^J(m_p)|^2 + \lambda \int_{{\bf M}_0} [L({\bf Y}_n^J)(m)]^2 dm \label{e:pen}
\end{align}
where $\tilde{{\bf \mathrm{y}}}_{np} = (\tilde{\mathrm{y}}_{np1}, \cdots, \tilde{\mathrm{y}}_{npD})^\top$, ${\bf Y}_n^J(m) = \sum_{j=1}^J b_{nj} {\bf e}_j(m)$, and $L$ is a linear differential operator. The resulting functional data would be 
\begin{equation}
\label{eq:fdobj1}
{\bf Y}_n(m) \approx \sum_{j=1}^J \hat{b}_{nj} {\bf e}_j(m). 
\end{equation}
In the ADAPT application we utilize felsplines \citep{ramsay:2002} and expand each coordinate, $Y_{nj}(m)$, separately, though other approaches including thin plate splines could also be used.

\subsection{2-Step Functional Principal Component Analysis}
\label{sec:method:fpca}

We now introduce a 2-step functional principal component analysis (FPCA) method to be carried out on the objects define in \eqref{eq:fdobj1}. In the first step, we conduct FPCA on the pooled (across the $D$ coordinates) sample to reduce the number of basis functions and make them orthogonal. In the second step, we conduct PCA on the resulting array to get eigenvalues $\lambda_k$ and eigenfunctions $\bV_k(m)$. Computational tools for basis functions that map a set in lower dimension ${\bf M}_0 \in \mathbb{R}^d$ to higher dimension, $\mathbb{R}^D$, are currently limited. Therefore, we start with expanding each coordinate of ${\bf Y}_n$ using basis functions $\{e_j: {\bf M}_0 \rightarrow \mathbb{R}\}$, and then obtain eigenfunctions ${\bf V}_k: {\bf M}_0 \rightarrow \mathbb{R}^D$ in the second step.

The raw data, $\{\mathrm{y}_{npq}\}$, is assumed to be an $N \times P \times D$ array, while the basis coefficients from \eqref{eq:fdobj1} form an $N \times J \times D$ array. Our second step consists of tensor multiplication and singular value decompositions, which are substantial computational burdens. Decreasing the dimension by lowering the number of basis functions through the first step lessens the computational time substantially. The burden of the second step is also decreased substantially by exploiting the orthonormal structure of the bases from the first step. 

\paragraph*{Step 1.} 
Without loss of generality, assume that the ${\bf Y}_n(m)$ have been centered and thus have mean zero. Each of the functions is expressed as
$$
{\bf Y}_{n}(m) = \left[ Y_{n1}(m), \cdots, Y_{nD}(m) \right]^\top \approx\left[
\sum_{j = 1}^J \hat b_{nj1} e_j(m), \cdots, \sum_{j = 1}^J \hat b_{njD} e_j(m)
\right]^\top,
$$
for $n = 1, \dots, N$. We stack all of the coordinate-wise functions into a single vector of functions with dimension $N D$. We denote the resulting functions as $Y_l(m)$ where $Y_l(m) = Y_{n q}(m)$ for $l = N(q-1)+n$ and $n = 1, \cdots, N$, $q = 1, \cdots, D$.

We find the pairs of eigenvalues $\eta_h$ and principal component functions $\psi_h: \bM_0 \to \mbR^D$, for $h=1,\dots, H$, which satisfy
\begin{gather}
\eta_h \psi_h(m) = \int \bPhi(m, m') \psi_h(m') dm', \label{eq:fpca1.1.1}\\
\|\psi_h\| = 1, \label{eq:fpca1.1.2}
\end{gather}
where
$$\bPhi(m,m') = E[Y_l(m)Y_l(m')^\top] = \sum_{j_1}^J \sum_{j_2}^J E[b_{lj_1} b_{lj_2}] e_{j_1}(m) e_{j_2}(m')^\top = \sum_{j_1}^J \sum_{j_2}^J \Pi_{j_1, j_2} e_{j_1}(m) e_{j_2}(m')^\top.$$
By expanding $\psi_h$ using $e_j$, $\psi_h(m) = \sum_{j=1}^J w_{hj} e_j(m)$, equations \ref{eq:fpca1.1.1} and \ref{eq:fpca1.1.2} become
\begin{gather*}
\eta_h \sum_{j=1}^J w_{hj} e_j(m) = \sum_{j_1}^J \sum_{j_2}^J \sum_{j_3}^J \Pi_{j_1, j_2} \bZ_{j_2, j_3} e_{j_1}(m), \label{eq:fpca1.2.1}\\
\text{and} \qquad
\sum_{j_1 = 1}^J \sum_{j_2 = 1}^J w_{hj_1} w_{hj_2} \bZ_{j_1, j_2} = 1, \label{eq:fpca1.2.2}
\end{gather*}
where $\bZ_{j_1, j_2} = \int_{{\bf M}_0} e_{j_1}(m') e_{j_2}(m') dm'$. By factoring the matrix $\bZ = \bG^\top \bG$ and defining $a_{hj} = \sum_{j_1=1}^J w_{h j_1} {\bf G}_{j j_1}$, we obtain the relations
\begin{gather*}
\eta_h a_{h j} = \sum_{j_2}^J \tilde{\Pi}_{j, j_2} a_{h j_2} 
\qquad \text{and} \qquad
\sum_{j=1}^J a_{hj}^2 = 1,
\end{gather*}
where $\tilde{\Pi}_{j, j_2} = \sum_{j_1}^J \sum_{j_3}^J {\bf G}_{j, j_1} \Pi_{j_1 j_3} {\bf G}_{j_3 j_2}$. Further details can be found in the supplementary material. The vector ${\bf a}_j = \{a_{h j}\}$ is the $j^{th}$ eigenvector of the covariance matrix $\tilde{\Pi}$. Using this, we can get $$\psi_h(m) = \sum_{j=1}^J a_{hj} e_j(m).$$

\paragraph*{Step 2.} We now expand $Y_{nq}(m)$ using the $\{\psi_h(m)\}$: $$Y_{nq}(m) = \sum_{h=1}^H c_{nhq} \psi_h(m).$$
The coefficients ${\bf c} = \{c_{nhq}\}$ form an $N \times H \times D$ array. The covariance operator of ${\bf Y}_n(m)$ is given by 
$$\Gamma_{q, q'} (m, m') = E[Y_{nq}(m) Y_{nq'}(m')] = \sum_{h_1=1}^H \sum_{h_2=1}^H \Sigma_{h_1 q h_2 q'} \psi_{h_1}(m) \psi_{h_2}(m')$$
with $\Sigma_{h_1 q h_2 q'} = E[c_{n h_1 q} c_{n h_2 q'}^\top].$  
Now we find the pairs of eigenvalues $\lambda_k$ and eigenfunctions $\bV_k(m)$ that satisfy
\begin{gather}
\lambda_k \bV_k(m) = \int \bGa(m,m') \bV_k(m') dm', \label{eq:fpca2.1.1}\\
\|\bV_k\|=1. \label{eq:fpca2.1.2}
\end{gather}
We expand $\bV_k$ using the $\psi_h$ as well: 
\begin{equation*} \label{eq:fpca2.0}
V_{kq}(m) = \sum_{h=1}^H v_{khq} \psi_h(m),
\end{equation*}
where ${\bf v} = \{v_{khq}\}$ is a $K \times H \times D$ array of coefficients. Since $\psi_h$ are orthonormal, after several expansions, which are given in the supplementary material, equations \ref{eq:fpca2.1.1} and \ref{eq:fpca2.1.2} become
\begin{gather*}
\lambda_k v_{k h q} = \sum_{q'=1}^D \sum_{h_2 = 1}^H \Sigma_{h q h_2 q'} v_{k h_2 q'}
\qquad \text{and} \qquad
\sum_{q=1}^D \sum_{h_1=1}^H \sum_{h_2=1}^H v_{k h_1 q} v_{k h_2 q}=1.
\end{gather*}
So we have that ${\bf v}_k = \{v_{k h q}\}$ for equation \ref{eq:fpca2.0} is the $k^{th}$ eigenmatrix of the $H \times D \times H \times D$ covariance tensor $\Sigma$.

\subsection{Manifold-on-Scalar Regression}
\label{sec:method:funreg}

We now give a Manifold-on-scalar regression strategy by using the functional manifold objects as responses and using scalar predictors, very similar to Function-on-Scalar Regression \citep{ramsay:silverman:2005}. The model is given by 
\begin{equation}
\label{eq:funreg}
{\bf Y}_n(m) = x_{n1} \bbe_{1}(m) + x_{n2} \bbe_{2}(m) + \cdots + x_{nR} \bbe_{R}(m) + \beg_n(m),
\end{equation}
where there are $R$ predictors for every manifold. Recall that $\bY_n$ is the deformation map associated with the manifold, $\mathcal{Y}_n$. Define the terms
\begin{align*}
& {\bf Y}(m) = \left[ \begin{matrix} \bY^\top_1(m)\\ \bY^\top_2(m)\\ \vdots\\ \bY^\top_N(m) \end{matrix} \right], \quad
{\bf X} = \left[ \begin{matrix} x_{11} & x_{12} & \cdots & x_{1R}\\ x_{21} & x_{22} & \cdots & x_{2R} \\ \vdots & \vdots & \vdots & \vdots\\ x_{N1} & x_{N2} & \cdots & x_{NR} \end{matrix} \right], \\
& {\bbe}(m) = \left[ \begin{matrix} \bbe^\top_1(m)\\ \bbe^\top_2(m)\\ \vdots\\ \bbe^\top_R(m) \end{matrix} \right], \quad
{\beg}(m) = \left[ \begin{matrix} \beg^\top_1(m)\\ \beg^\top_2(m)\\ \vdots\\ \beg^\top_N(m) \end{matrix} \right],
\end{align*}
then equation \ref{eq:funreg} can be expressed as
$${\bf Y}(m) = {\bf X}\boldsymbol{\beta}(m) + \beg(m).$$
The least square estimator of the functional parameter $\boldsymbol{\beta}(m)$ can be found by minimizing
\begin{equation} \label{eq:funreg:obj}
\sum_{n=1}^N \left\| \bY_n - \sum_{r=1}^R x_{nr}\bbe_{r} \right\|^2 = \int_{{\bf M}_0} \sum_{n=1}^N \left| \bY_n(m) - \sum_{r=1}^R x_{nr}\bbe_{r}(m) \right|^2 dm
\end{equation}
For each fixed $m$, $\int_{{\bf M}_0} \sum_{n=1}^N e_n^2 (\boldsymbol{\beta}, m) dm$ is minimized if
\begin{equation}
\label{eq:betahat}
\boldsymbol{\hat{\beta}}(m) = ({\bf X}^\top {\bf X})^{-1} {\bf X}^\top {\bf Y}(m).
\end{equation}
We expand $\bY_n(m)$ using functional principal components, ${\bf V}_k(m) \in \mathbb{R}^{3}$, from Section \ref{sec:method:fpca}:
$$\bY(m) = \sum_{k=1}^K y_{nk} {\bf V}_k(m).$$
Let 
$${\bf Y}(m) 
= \left[ \begin{matrix} y_{11} & y_{12} & \cdots & y_{1K}\\ y_{21} & y_{22} & \cdots & y_{2K} \\ \vdots & \vdots & \vdots & \vdots\\ y_{N1} & y_{N2} & \cdots & y_{NK} \end{matrix} \right] \left[ \begin{matrix} {\bf V}^\top_1(m)\\{\bf V}^\top_2(m)\\ \vdots\\ {\bf V}^\top_K(m) \end{matrix} \right] 
\triangleq {\bf y}{\bf V}(m).
$$
Then 
$$\boldsymbol{\hat{\beta}}(m) = ({\bf X}^\top {\bf X})^{-1} {\bf X}^\top {\bf y}{\bf V}(m).$$

While least squares works well, it can often be improved by penalizing the roughness of the resulting $\hat\beta$'s.  In this case the objective function \ref{eq:funreg:obj} changes to
\begin{equation} \label{eq:funreg:penalobj1}
\sum_{n=1}^N \int_{{\bf M}_0} \left| {\bf Y}_n (m) - \sum_{r=1}^R x_{nr}  \boldsymbol{\beta}_r (m) \right|^2 dm + \sum_{r=1}^R \lambda_r \int_{{\bf M}_0} |L \boldsymbol{\beta}_r (m)|^2 dm,
\end{equation}
where $\lambda_r$ is tuning parameter and $L$ is a roughness operator. It is important to choose $L$ carefully. Since we do not want the minimizer of \ref{eq:funreg:penalobj1} to change based on the coordinate system, we need an operator that is invariant to rotation and translation. For this reason \citet{ramsay:2002} chose the Laplacian operator. In the case where ${\bf f}: [{\bf M}_0 \subset \mathbb{R}^2] \rightarrow \mathbb{R}^3$,  the Laplacian operator of ${\bf f}$ is given by
$$\bigtriangleup {\bf f} = \bigtriangleup [f_1, f_2, f_3]^\top = [\bigtriangleup f_1, \bigtriangleup f_2, \bigtriangleup f_3]^\top = \left[\frac{d^2 f_1}{d m_1^2} + \frac{d^2 f_1}{d m_2^2}, \frac{d^2 f_2}{d m_1^2} + \frac{d^2 f_2}{d m_2^2}, \frac{d^2 f_3}{d m_1^2} + \frac{d^2 f_3}{d m_2^2} \right]^\top,$$ 
where $f_1, f_2, f_3$ correspond to each coordinate of ${\bf f}$ and $m_1$ and $m_2$ correspond to each coordinate of ${\bf M}_0$.
${\bf Y}_n$ and ${\bf \beta}_r$ can both be expanded with PC functions ${\bf V}_k: [{\bf M}_0 \subset \mathbb{R}^2] \rightarrow \mathbb{R}^3$ for $k = 1, \cdots, K$:
\begin{equation}
\label{eq:funreg:penalobj2}
\sum_{n=1}^N \int_{{\bf M}_0} \left| \sum_{k=1}^K y_{nk} {\bf V}_k (m) - \sum_{r=1}^R x_{nr} \sum_{k=1}^K b_{rk} {\bf V}_k (m) \right|^2 dm + \sum_{r=1}^R \lambda_r \int_{{\bf M}_0} \left| \sum_{k=1}^K b_{rk}  ( \bigtriangleup {\bf V}_k (m) ) \right|^2 dm.
\end{equation}
Let $$B = \left[ \begin{matrix} b_{11} & b_{12} & \cdots & b_{1K}\\ b_{21} & b_{22} & \cdots & b_{2K} \\ \vdots & \vdots & \vdots & \vdots\\ b_{R1} & b_{R2} & \cdots & b_{RK} \end{matrix} \right], \quad 
\Lambda = \left[ \begin{matrix} \lambda_1 & 0 & \cdots & 0\\ 0 & \lambda_2 & \cdots & 0 \\ \vdots & \vdots & \vdots & \vdots\\ 0 & 0 & \cdots & \lambda_R \end{matrix} \right].$$
Then objective \ref{eq:funreg:penalobj2} becomes to find $B$ that minimizes
\begin{equation*}
trace\{({\bf y} - {\bf X} B)^\top ({\bf y} - {\bf X} B)\} + trace\{\Lambda B U B^\top \},
\end{equation*}
where 
$$U_{{k_1}, {k_2}} = \int_{{\bf M}_0} ( \bigtriangleup {\bf V}_{k_1} )^\top ( \bigtriangleup {\bf V}_{k_2} ) dm,$$
and the least square estimate of $B$ is
\begin{equation*}
\text{vec}(\hat{B}^\top) = \left( ({\bf X}^\top {\bf X}) \otimes I_K + \Lambda \otimes U^\top \right)^{-1} \text{vec}({\bf y}^\top {\bf X}).
\end{equation*}
See supplement for further details.\\

To test for the significance of $\boldsymbol{\hat{\beta}}$, we find the asymptotic distribution of $\boldsymbol{\hat{\beta}}$.
Assume that 
$$Y_n(m) = {\bf X}_n^\top \boldsymbol{\beta}(m) + \epsilon_n (m),$$ 
where $\{{\bf X}_n\}$ are iid random elements of $\mathbb{R}^R$ whose covariance matrix, $\Sigma_{\bf X}$, exists and has full rank. Also assume that $\{\epsilon_n\}$ are mean zero iid elements of $L^2[{\bf M}_0]$ with $E\|\epsilon_n\|^2 < \infty$, which implies the covariance function, ${\bf C}_\epsilon(m,m')$, of $\epsilon_n(m)$ exists. Assume that the sequences $\{{\bf X}_n\}$ and $\{\epsilon_n\}$ are independent of each other. Then we have by the CLT for Hilbert spaces that
$$\sqrt{N}(\boldsymbol{\hat{\beta}} - \boldsymbol{\beta}) \xrightarrow[]{d} \mathcal{N}(0, {\bf C}_\beta),$$
where ${\bf C}_\beta (m,m')$ is an $R \times D \times R \times D$ array at each pair $(m, m') \in \mathcal{M}_0 \times \mathcal{M}_0$ and equals
$${ C}_{i,j,k,l;\beta} (m,m') = (\Sigma^{-1})_{i,k;\bf X}  C_{j,l;\epsilon}(m,m').$$

The covariance of the errors can be estimated as 
$${\bf \hat{C}}_\epsilon (m, m') = \frac{1}{N-R}\sum_{n=1}^N (\bY_n(m) - {\bf X}_n^\top \boldsymbol{\hat{\beta}}(m)) (\bY_n(m') - {\bf X}_n^\top \boldsymbol{\hat{\beta}}(m'))^\top. $$
Notice this ${\bf \hat{C}}_\epsilon$ corresponds with a $P \times D \times P \times D$ array because at every $(m_p, m_{p'})$, ${\bf \hat{C}}_\epsilon (m_p, m_{p'})$ is $D \times D$ matrix, and there are $P$ points $m_p$. 
For each $\beta_r(m)$,
$$\sqrt{N}({\hat{\beta}_r} - \beta_r) \xrightarrow[]{d} \mathcal{N}(0, {\bf C}^r_\beta),$$
and we estimate 
$${\bf \hat{C}}^r_\beta (m,m')= N({\bf X}^\top {\bf X})^{-1}_{r,r} {\bf \hat{C}}_\epsilon (m, m').$$

We now test the significance of $\hat{\beta}_r$. We can do this pointwise, i.e. test $\hat{\beta}_r(m_p)$ at each point, and we can also find the overall confidence region around the face using the strategy of \citet{choi:reimherr:2016}.  We call this simultaneous confidence region a \textit{confidence bubble} as it forms a 3D region around the parameter estimates.  \\
We first rotate $\hat{\beta}_r$ and get $\hat{\beta}'_r = ({\bf \hat{C}}^r_\beta (m,m))^{-1/2}\hat{\beta}_r$. Then
$$\sqrt{N}(\hat{\beta}'_r - \beta'_r) \xrightarrow[]{d} \mathcal{N}(0, {\bf \tilde{C}}^r_\beta)$$
where
$${\bf \tilde{C}}^r_\beta (m,m') = ({\bf \hat{C}}^r_\beta (m,m))^{-1/2} {\bf \hat{C}}^r_\beta (m,m') ({\bf \hat{C}}^r_\beta (m',m'))^{-1/2}.$$
Pointwise we have
$$\sqrt{N}(\hat{\beta}'_r (m_p) - \beta'_r (m_p)) \xrightarrow[]{d} \mathcal{N}(0, I_{3\times3}).$$
Therefore, $$T_{norm}^{pt} = N\|(\hat{\beta}'_r (m_p) - \beta'_r (m_p)\|^2 \xrightarrow[]{d} \chi^2 (3).$$\\

We conclude by providing a strategy for constructing simultaneous confidence ellipses for each $\hat \beta(m)$, which is based on a technique from \citet{choi:reimherr:2016}.  The proof can be found in the online supplemental.
\begin{theorem}\label{t:bubble}
If $\sqrt{N}(\hat{\beta}_r - \beta_r)$ converges in distribution to a Gaussian process, $\mathcal{N}(0, {\bf C}^r_\beta)$, and the square-root of the eigenvalues, $\{\lambda_i\}$, of ${\bf C}^r_\beta$ are summable, then 
$$
P \left\{ \sqrt{N}|\hat{\beta}_r(m) - \beta_r(m)| \leq \sqrt{\xi_\alpha \sum_{j=1}^\infty \sqrt{\lambda_j} |{\bf U}_j(m)|^2}, \ \text{for almost all }m \in \bM_0 \right\}
\leq \alpha + o(1),
$$
where $\{ {\bf U}_j\}$ are the eigenfunctions of ${\bf C}^r_\beta$, and $\xi_\alpha$ is such that $P(\sum_{j=1}^\infty \sqrt{\lambda_j} Z_j^2 > \xi_\alpha) \xrightarrow[]{d} \alpha$.
\end{theorem}

\section{Simulation Studies}
\label{sec:simulation}

To compare the performance of manifold-on-scalar regression to the performance of multivariate principal component regression, PCR, we consider the following simulation setting. We construct simulated faces using the following model:
\begin{equation*} \label{eq:sim_model}
Y_n(m_{ni}) = \delta \times X \times \beta (m_{ni}) + \epsilon (m_{ni}) + \gamma_{ni}.
\end{equation*}
Here $\delta$ is a positive constant signifying the strength of the effect, $X \sim \mathcal{N}(0,1)$, $\beta (m)$ is a coefficient function, $\epsilon (m)$ is an error function, and $\gamma_{ni}$ is an iid measurement noise. To ensure a realistic simulation, we take $\beta(m)$ to be the estimated $\beta (m)$ for height from Section \ref{sec:adapt}. A plot of this $\beta$ is given in the top left of Figure \ref{fig:sim_betahat}. The error, $\varepsilon(m)$ is constructed by randomly selecting one of the faces from the ADAPT study, while the $\gamma_{ni}$ is a vector of $\mathcal{N}(0,0.002)$ displacements to each x, y, and z coordinate.
The examples of simulated faces are as in Figure \ref{fig:sim_examples}. The plots shown are based on $\delta = 20$.

\begin{figure}[ht]
\begin{center}
\includegraphics[width = .95\linewidth]{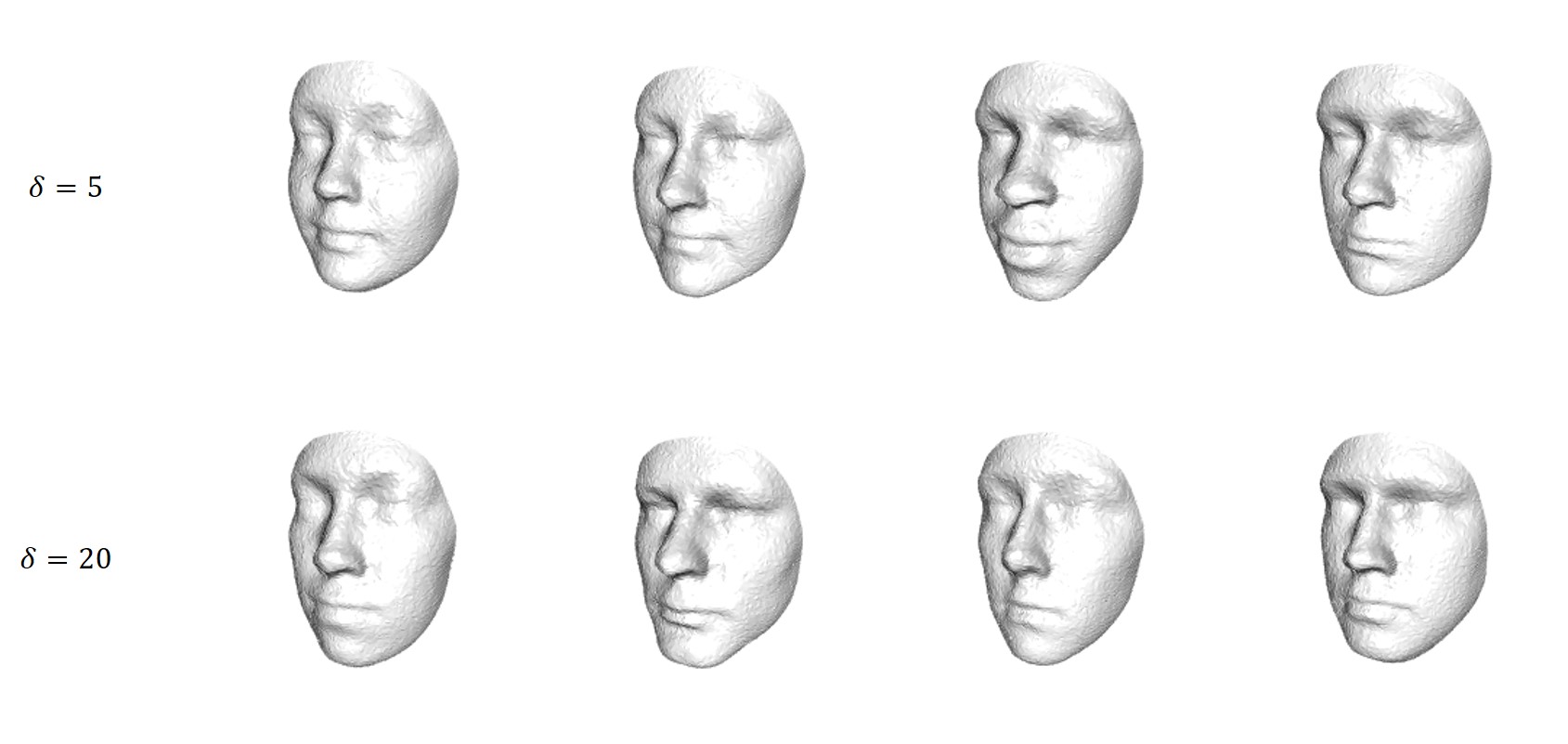}
\caption{Examples of simulated faces for $\delta = 5$ (top) and $\delta = 20$ (bottom).}
\label{fig:sim_examples}
\end{center}
\end{figure}

We repeated the simulation 1000 times for $\delta = 0$, $\delta = 5$, $\delta = 20$, $\delta = 50$, $\delta = 100$, and $\delta = 200$ and for each run $N=100$ is taken. We compared the rejection rates based on PCA test, Choi test, and norm test for multivariate PCR and functional PCR. For multivariate PCR, we ran principal component analysis on pooled data, stacking x-coordinate values, y-coordinate values, and z-coordinate values of the 7150 quasi-landmarks on the faces. We then took principal components that explain 99 \% of the total variation, used them as our response, and fit a linear regression model with predictor $X$. We have compared this to our method, functional PCR. 
It is not very conventional to conduct a PCA test, Choi test, or norm test in the multivariate case, since those tests target infinite dimensional spaces. However, to make a proper comparison, we have used the same testing methods.

The results are summarized in Table \ref{table:sim_smooth}. It shows that the rejection rates for $\delta = 0$ is within 2 standard error of 0.05, the alpha level we took. As $\delta$ increases, the rejection rate increases as expected, and when $\delta = 200$ the rejection rate becomes almost 1. In most of cases, the rejection rates for functional PCR are bigger than the rejection rates for multivariate PCR, except a few cases like Choi test for $\delta=20$.

\begin{table}[ht]
\begin{center}
\begin{tabular}{| c | c | c | c | c |}
\hline
$\delta$ & type & rej rate PCA & rej rate Choi & rej rate norm\\ \hline
   0  & multivartate & 0.044 & 0.061 & 0.038\\
       & functional     & 0.059 & 0.064 & 0.060\\ \hline
  5   & multivariate & 0.054 & 0.070 & 0.049\\
       & functional     & 0.078 & 0.078 & 0.069\\ \hline
 20  & multivariate & 0.063 & 0.090 & 0.061\\
       & functional     & 0.097 & 0.101 & 0.094\\ \hline
 50  & multivariate & 0.136 & 0.221 & 0.181\\
       & functional     & 0.265 & 0.299 & 0.293\\ \hline
100 & multivariate & 0.474 & 0.739 & 0.596\\
       & functional     & 0.662 & 0.768 & 0.745\\ \hline
200 & multivariate & 0.989 & 1.000 & 0.997\\
       & functional     & 0.999 & 1.000 & 0.999\\ \hline
\end{tabular}
\label{table:sim_smooth}
\caption{The rejection rates based on three different tests (PCA test, Choi test, and norm test) for different $\delta$'s. For $\delta=0$ case, the rejection rates are approximately 0.05, the alpha in this case, and for the other cases, the rejection rates for functional PCR are higher than the rejection rates for multivariate PCR.}
\end{center}
\end{table}

Some examples of estimated betas from functional PCR and from multivariate PCR are given in Figure \ref{fig:sim_betahat}. Red and yellow colors are where beta shows outward effect, meaning that in those parts the face goes outward when predictor increases, while blue and skyblue means that beta shows inward effect. Since it is height we have used, the plot in left shows that the face would become prolonged as the predictor increases. And the plots show that the estimated beta from functional PCR picks up the smoothness of the original beta and better resembles the original beta, while the estimated beta from multivariate PCR shows rough edges and sometimes gives very different effect as in the bottom ($\delta=20$ case).

\begin{figure}[ht]
\begin{center}
\includegraphics[width = .65\linewidth]{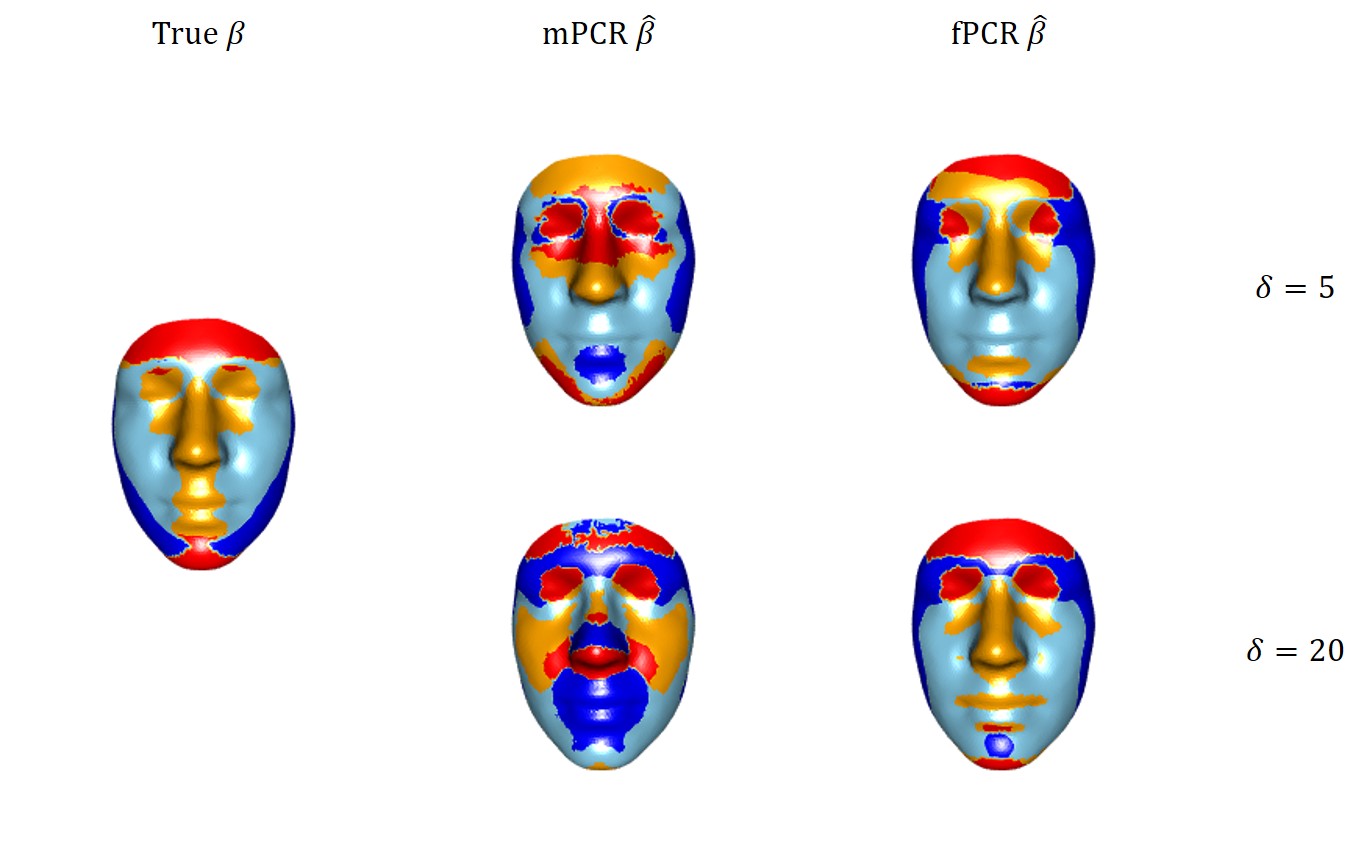}
\caption{Examples of estimated beta. The left is the beta used for simulation, the middle is the estimated beta from multivariate PCR, and the right is the estimated beta from functional PCR. The top row is for $\delta = 5$ and the bottom row is for $\delta = 20$. Red and yellow means that beta shows outward effect while blue and skyblue means that beta shows inward effect.}
\label{fig:sim_betahat}
\end{center}
\end{figure}

\section{ADAPT Study}
\label{sec:adapt}

This section presents the application of our methodologies from Section \ref{sec:method} to the ADAPT data. We convert the 3D facial imaging data into functional objects in Section \ref{sec:adapt:fdobj}, where we also discuss the details on how to apply each step of the framework in Section \ref{sec:method:fdobj}. Section \ref{sec:adapt:fpca} presents the principal components of our 2-step FPCA from Section \ref{sec:method:fpca}. Section \ref{sec:adapt:funreg} presents a regression model with the 3D faces as manifold outcomes and the covariates age, gender, height, weight, and genetic ancestry; we discuss the effects and significances of the resulting coefficient functions.

\subsection{Facial Functional Object Construction}
\label{sec:adapt:fdobj}

We view each face as a 2-dimensional manifold that is a subset of $\mathbb{R}^3$, and our goal is to construct functional objects ${\bf Y}_n: {\bf M}_0 \rightarrow \mathbb{R}^3$ from each face. There are 6564 faces, and each face is sampled densely with 7150 points in x, y, and z coordinates. Therefore, the data is $\{ \mathrm{y}_{npq}: n = 1, \cdots, 6564; p = 1, \cdots, 7150; q = 1, 2, 3 \}$. We elaborate each step of constructing facial functional objects as below. \\

\noindent {\bf Step 1. } We identified a reference face $\mathcal{M}_0$ as the mean of the 6564 faces, that is, $\{\bar{\mathrm{y}}_{pq}: \bar{\mathrm{y}}_{pq} = \frac{1}{N}\sum_{n=1}^N \mathrm{y}_{npq}; p = 1, \cdots, 7150; q = 1, 2, 3\}$.  This approach is possible because the data were already aligned via Procrustes analysis; the resulting mean face is given in Figure \ref{fig:whichinner}.\\

\begin{figure}[ht]
\begin{center}
\includegraphics[width = .4\linewidth]{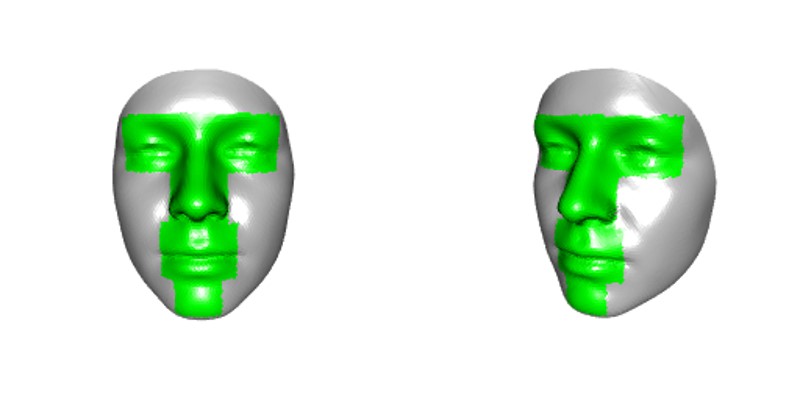}
\caption{Plots of mean face which is taken as a reference face. Green area represents the area where finer mesh is taken for FELSPLINE basis functions. Examples of mesh plots are as in Figure \ref{fig:mani_mesh}.}
\label{fig:whichinner}
\end{center}
\end{figure}

\noindent {\bf Step 2. } We apply manifold learning techniques to the mean face to find ${\bf M}_0$, the representation of mean face in $\mathbb{R}^2$. The resulting ${\bf M}_0$ is represented by $\{\mathrm{m}_{pq}; p = 1, \cdots, 7150, q = 1, 2, 3\}$. 
The choice of the manifold learning technique for constructing ${\bf M}_0$ is important to obtain a reasonable functional object that is close to the original data. Figure \ref{fig:mani_mesh} shows how ${\bf M}_0$ changes with different manifold learning techniques.
Since smoothness is defined with distance along $\mathcal{M}_0$, we believed that the manifold learning techniques that preserve local distances would work best. In order to check our intuition, we tried several nonlinear dimension reduction techniques like local linear embedding (LLE, \citet{saul:roweis:2003}), Laplacian eigenmaps \citep{belkin:niyogi:2003}, 
Isomap \citep{tenenbaum:desilva:langford:2000}, local tangent space alignment (LTSA, \citet{zhang:zha:2004}), Diffusion Map \citep{nadler:lafon:coifman:kevrekidis:2006}, and Spanifold \citet{chenouri:2015:spanifold} along with a linear dimension reduction technique principal component analysis (PCA) for a comparison.\\

\noindent {\bf Step 3. } We need to align all faces to the reference face, but the faces are already aligned using a generalized Procrustes superimposition \citep{rohlf:slice:1990} as discussed in Section \ref{sec:intro:data}. Therefore, we can infer that the faces are also aligned to the reference manifold $\mathcal{M}_0$, the mean face. \\

\noindent {\bf Step 4. } We construct basis functions ${\bf e}_j: [{\bf M}_0 \subset \mathbb{R}^2] \rightarrow \mathbb{R}^3$, but given the limitations in constructing such basis functions, we took basis functions $e_j: [{\bf M}_0 \subset \mathbb{R}^2] \rightarrow \mathbb{R}$ to expand the functional objects marginally
\begin{equation}
\label{eq:facialobj}
{\bf Y}_n(m) = \begin{bmatrix} Y_{n1} \\ Y_{n2}\\ Y_{n3} \end{bmatrix} (m) = \sum_{j=1}^J \begin{bmatrix} b_{nj1}\\ b_{nj2}\\ b_{nj3} \end{bmatrix} e_j (m)
\end{equation}
\noindent where $Y_{n1}$ corresponds to x coordinate of ${\bf Y}_n$, $Y_{n2}$ corresponds to y coordinate of ${\bf Y}_n$, and $Y_{n3}$ corresponds to z coordinate of ${\bf Y}_n$. We find $\{\hat{b}_{njq}\}$ for $q = 1,2,3$ by minimizing equation \ref{e:pen}.
We used FELSPLINEs \citep{ramsay:2002} which are designed for irregularly shaped domain with complex boundaries and use a finite element method, meaning that the domain is divided into triangular meshes and piecewise linear and quadratic functions are fit on each mesh. Therefore, we needed to create meshes out of our domain. \citet{ramsay:2002} uses all data points as vertices of the mesh, but in our case that will return over twenty thousand basis functions. In order to limit the number of basis functions to less than the number of observations per face, which is 7150 in our data, we created new meshes using the {\tt R} package \texttt{INLA} \citep{lindgren:rue:2013}. \\

There can be many different ways to create meshes, and the choice of mesh is closely related to the number of basis functions, thus affecting how close the functional objects are to the data. We took a finer mesh around periorbital, perinasal, and perioral areas shown as the green area in Figure \ref{fig:whichinner}, as these localized facial features are emphasized in \citep{hammond:et:al:2005}, and a coarser mesh around the cheeks and forehead where the surface is more smooth. The meshes for different manifold learning techniques are given in the bottom row  of plots of Figure \ref{fig:mani_mesh}. 

\begin{figure}[ht]
\begin{center}
\hspace{.5mm}
\includegraphics[width = .95\linewidth]{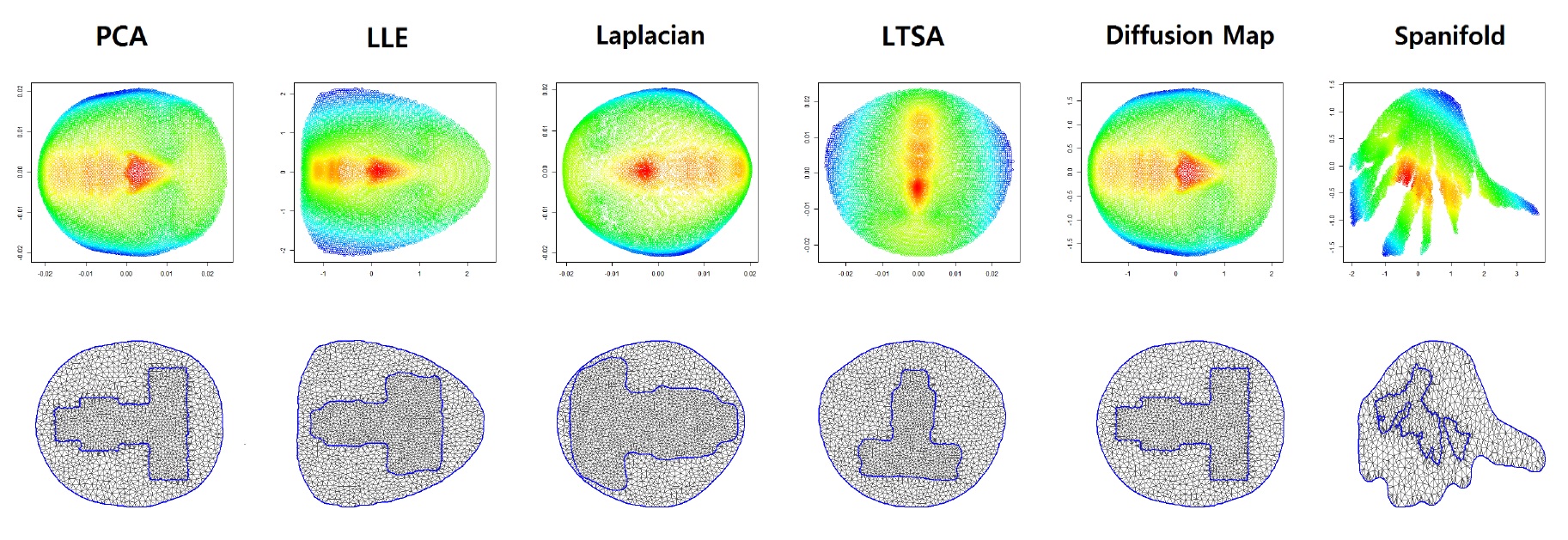}
\caption{The dimension-reduced reference manifold ${\bf M}_0$ and corresponding mesh using different manifold learning techniques. The finer inner mesh correspond to the area of green in Figure \ref{fig:whichinner}.}
\label{fig:mani_mesh}
\end{center}
\end{figure}

Table \ref{table:compare_maniAMSE} presents the average mean squared errors (AMSE) of 100 randomly selected faces, which is a measure of how close the functional objects ${\bf Y}_n(m)$ are to the original data $\{\mathrm{y}_{npq}\}$:
\begin{equation}
\label{eq:amse}
\text{AMSE:} \quad \quad \frac{1}{N} \frac{1}{P}\sum_{n=1}^{N} \sum_{p=1}^{P} | {\bf \mathrm{y}}_{np} - {\bf Y}_n(m_p) |^2,
\end{equation}
\noindent where $N = 100$ and $P=7150$.  We stress that, at this stage, we are not aiming for dimension reduction; our goal is to approximate the data using basis functions with as little error as possible.  Therefore, in this step we want the AMSE to be as small as possible to minimize any information loss when converting to functional objects.  However, the Procustes Analysis used to initially align and scale the data results in a unit-less scale for the coordinates of the face; the x-axis has been rescaled to have a range of 1.  This means that the AMSE values themselves are difficult to interpret, and thus we focus on comparisons of the AMSE's.  
The range of the first coordinate of the domain points ${\bf m}_n$, $\{\mathrm{m}_{p1}\}$, is different for each ${\bf M}_0$ from the different manifold learning techniques, and thus we made the smoothing parameter, $\lambda$, in equation \ref{e:pen} dependent on the range of x. The AMSE for LTSA with $\lambda_3$ is smallest, while the AMSE for LLE with $\lambda_1$ is also similarly small. Both LTSA and LLE try to preserve neighborhood distances of the original manifold, which confirms our intuition that they would best represent smoothness defined with distances along ${\bf M}_0$ and thus give a good fit. Spanifold gives the largest AMSE, which is not surprising given that the ${\bf M}_0$ is very irregular. This is because a human face face has many local peaks, and Spanifold works better with more regular surfaces.

\begin{table}[ht]
\begin{center}
\begin{tabular}{| c | c | c | c | c | c | c |}
\hline
& PCA & LLE & Laplacian & LTSA & Diffusion Map & Spanifold \\ \hline
$\lambda_1$ & 0.00247 & 0.00018 & 0.00423 & 0.00078 & 0.00056 & 0.00679
\\
$\lambda_2$ & 0.00084 & 0.00036 & 0.00167 & 0.00017 & 0.00078 & 0.00690 \\
$\lambda_3$ & 0.00058 & 0.00083 & 0.00114 & {\bf 0.00013} & 0.00142 & 0.00772\\ \hline
range(x) & 0.046 & 3.900 & 0.041 & 0.051 & 3.906 & 5.671\\ \hline
\end{tabular}
\label{table:compare_maniAMSE}
\caption{The pointwise mean squared errors of ${\bf Y}(m)$ of 100 randomly selected faces as in equation \ref{eq:amse} for different $\lambda$'s from mesh as in Figure \ref{fig:mani_mesh}. $\lambda_1$ = range(x)/$10^4$, $\lambda_2$ = range(x)/$10^5$, $\lambda_3$ = range(x)/$10^6$ where range(x) is the range of $\{\mathrm{m}_{p1}\}$.}
\end{center}
\end{table}

For all subsequent analyses, we utilize the manifold  objects constructed using the presented LTSA mesh and used $\lambda$ a little less than $\lambda_3$ to recover the details of face. Figure \ref{fig:faces} shows that the facial functional objects are very close to the original faces except for some smoothing. Figure \ref{fig:PMSE} shows a heatmap of the pointwise errors between functional objects and the original data. The tip of the nose shows a  relatively high pointwise error compared to the other areas, which is due to smoothing.  The boundary does not show much deviation and seems to be stable. We believe the resulting objects are reasonable approximations of the original faces.

\begin{figure}
\centering
\includegraphics[width = 0.45\linewidth]{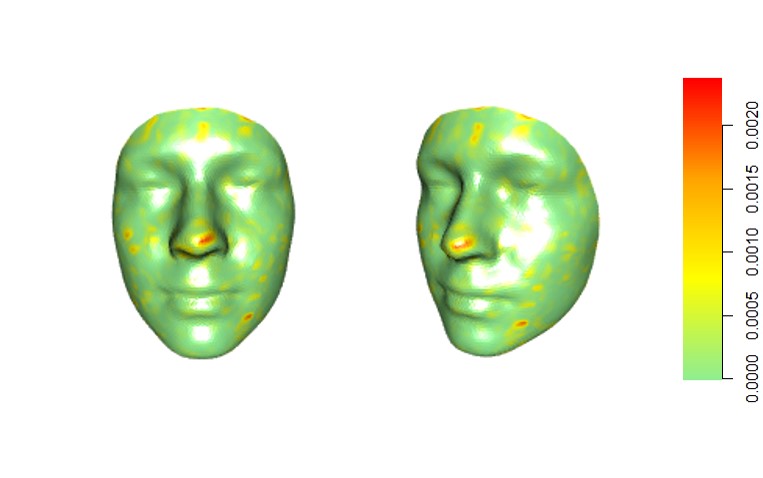}
\caption{Plot shows pointwise mean squared errors across all 6564 faces. This shows that the difference between the original faces and facial functional objects are very small.} 
\label{fig:PMSE}
\end{figure}

\begin{figure}[ht]
\centering
\includegraphics[width = .7\linewidth]{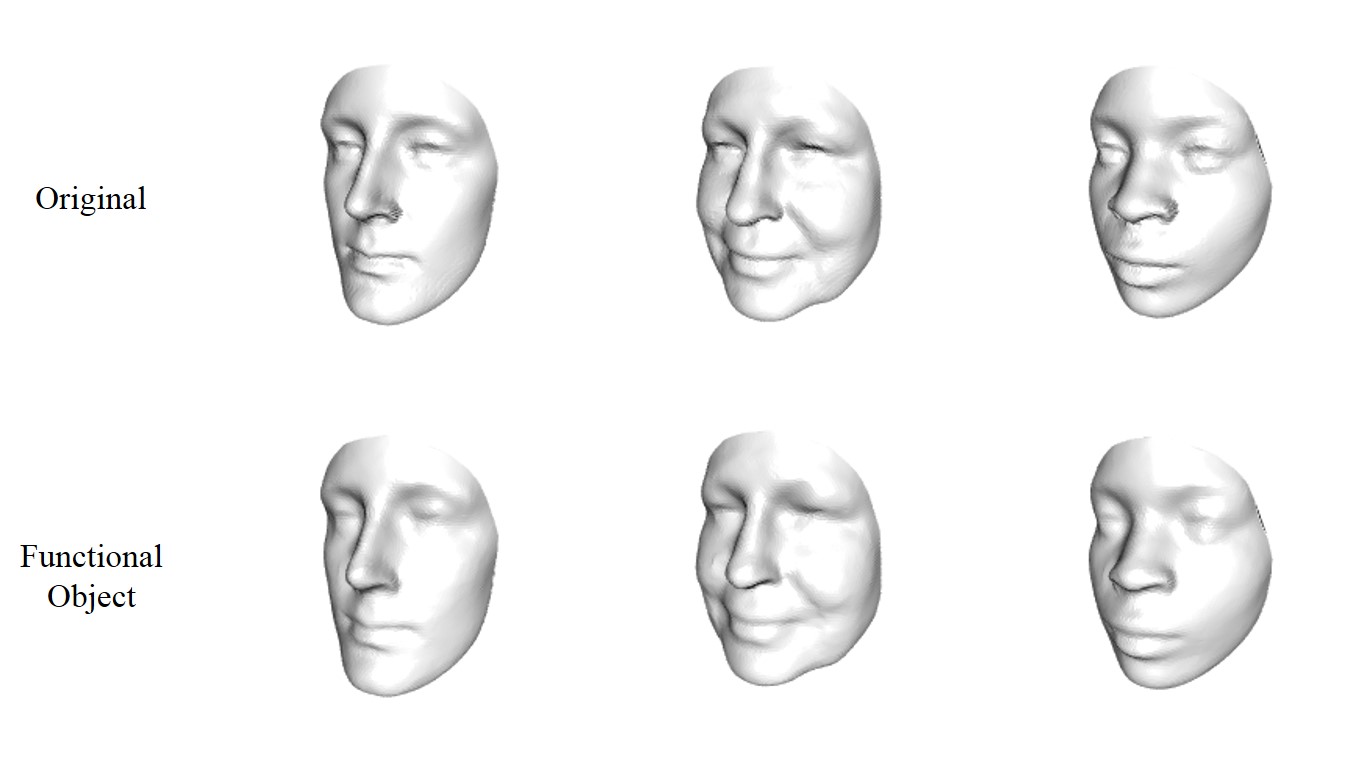}
\caption{Top three plots are examples of facial data of ADAPT, $\{\mathrm{y}_{npq}\}$, and bottom three plots are corresponding facial functional objects, ${\bf Y}_n(m)$. This shows that the facial functional objects closely resembles the original faces.}
\label{fig:faces}
\end{figure}

\subsection{Functional Principal Component Analysis}\label{sec:adapt:fpca}

In this section we apply the 2-step Functional Principal Component Analysis (2-step FPCA) discussed in Section \ref{sec:method:fpca} to the ADAPT data. We take $H=200$ principal components, or $\psi_h(m)$, in the first step (pooling coordinates), which accounts for 99.9\% of the total variance.  In the second step we then compute the PCs without pooling coordinates,$\bV_k(m)$, and Figure \ref{fig:cumvarprop} shows the cumulative proportion of explained variance. The first principal component $\bV_1(m)$ explains 31.27\%, the second principal component $\bV_2(m)$ explains 12.43\%, and the third principal component $\bV_3(m)$ explains 10.59\% of variation. The first 5 principal components combined explain 66.71\%, and the first 10 principal components combined explain 81.26\%.

\begin{figure}[ht]
\begin{center}
\includegraphics[width = .6\linewidth]{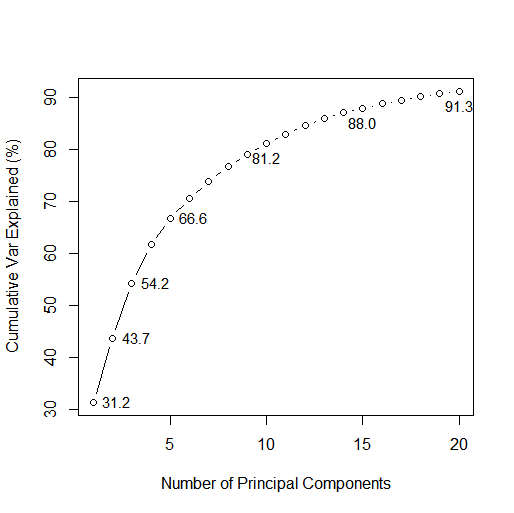}
\caption{Cumulative proportion of variance for number of PCs. First 10 PCs explain about 81.2\% of total variance and first 18 PCs explain about 90.2\% of total variance. }
\label{fig:cumvarprop}
\end{center}
\end{figure}

In Figure \ref{fig:pc_dir}, we demonstrate how each principal component affects the face, which, given that we are working in 3D is a bit challenging to visualize.  We thus compute the orthogonal vector, ${\bf t}_p$, to the tangent plane of each facial point, ${\bf \mathrm{m}}_p$, by conducting traditional PCA in a small neighborhood of $m_p$ (distance 0.1). As the first and second principal components would be the two vectors spanning the tangent plane, the third principal component would be the vector orthogonal to the tangent plane. Note that PCA also gives $|{\bf t}_p|^2 = 1$. We then calculated the inner product $\langle \bV_k(m_p), {\bf t}_p \rangle$ at each point for $p=1, \cdots, 7150$. The yellow to red area in Figure \ref{fig:pc_dir} denotes a PC whose effect points outward while the lightblue to blue area means that the effect of PC at that point is inward. Orange and lightblue mean weaker effects and red and blue mean stronger effects. 

\begin{figure}[ht]
\begin{center}
\includegraphics[width = .95\linewidth]{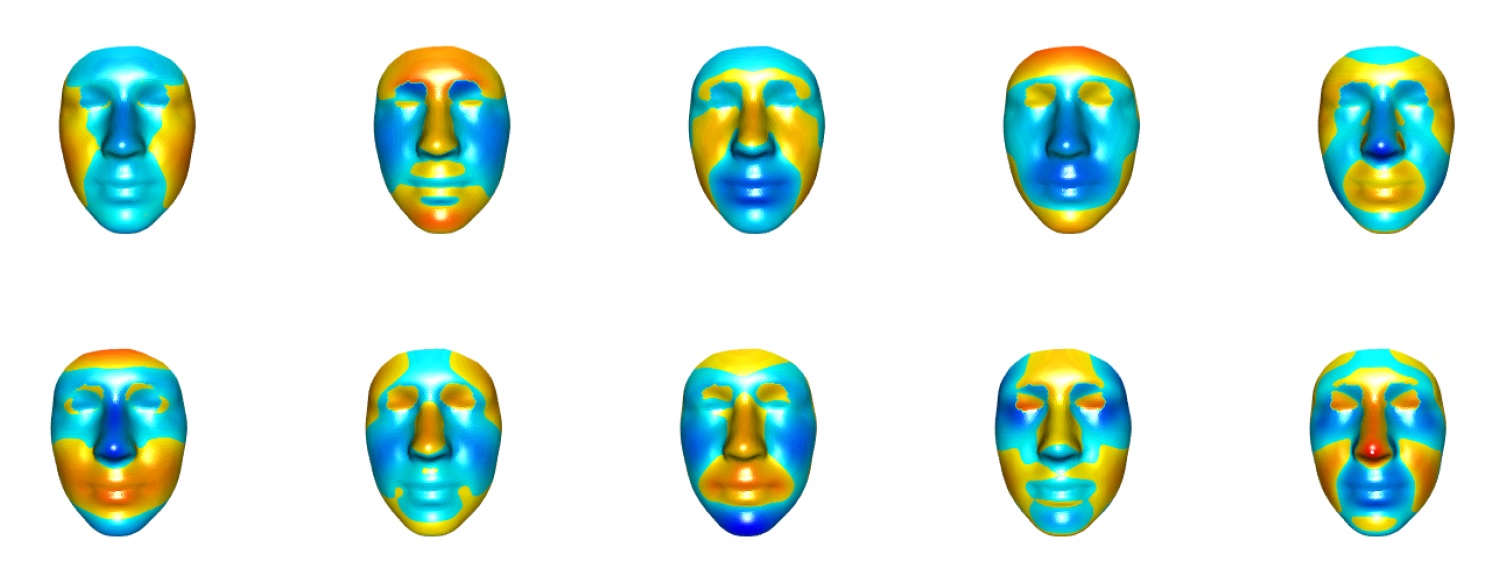}
\caption{The directional plots for PC 1-5 on the top and PC 6-10 on the bottom. The color on the face shows the direction and the strength, from weakest to strongest, of each PC effect on face: from lightblue to blue, inward, and from yellow to red, outward.}
\label{fig:pc_dir}
\end{center}
\end{figure}

As the top leftmost plot of Figure \ref{fig:pc_dir} suggests, the major difference between the mean face and the reconstructed faces using the first PC in Figure \ref{fig:pc_faces} is the sides of the faces. The top face became thinner while the middle face became a little rounder on the cheek. Figure \ref{fig:pc_faces} shows the progression of facial changes with more PCs included.  The rightmost faces are good approximations to the bottom plots in Figure \ref{fig:faces}, explaining 91.39\% of total variation. Thus we have now reduced the dimension of the data from 7150 points to 20 principal components, while carefully controlling the information loss.

\begin{figure}[ht]
\begin{center}
\includegraphics[width = .95\linewidth]{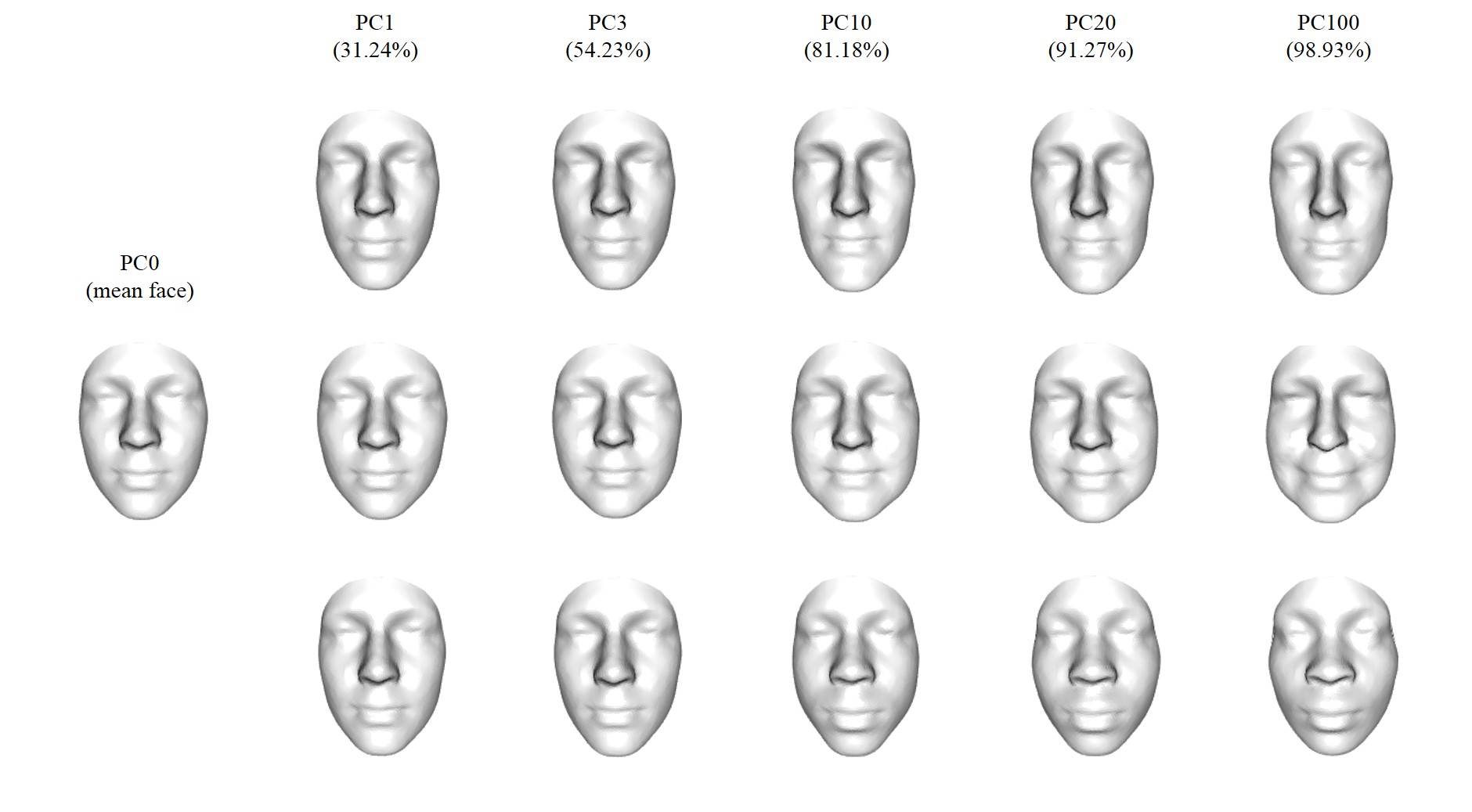}
\caption{Three facial functional objects expanded using different number of PCs from the second step of FPCA. Leftmost plot is the mean face. 
The percentage of variation explained is given at the top of each column.}
\label{fig:pc_faces}
\end{center}
\end{figure}

\subsection{Manifold-on-Scalar Regression}
\label{sec:adapt:funreg}

We conclude the application section by carrying out Manifold-on-Scalar Regression, which represents a major strength of our methodology.  
We examine the effects of sex, age, height, weight, and genetic ancestry the structure of human faces.  Genetic ancestry is measured as a proportion of a particular ethnic background, where E.ASN refers to East Asian, S.ASN refers to South Asian, AMR refers to Native American, W.AFR refers to West African, and S.EUR refers Southern European. There is also N.EUR which refers to Northern European, but since the sum of all proportions is 1, it is removed from the covariates, meaning that it is acting as the ancestral baseline, so all ancestral effects indicate differences from Northern Europeans. For the response variable, we take the facial functional objects ${\bf Y}_n(m)$ expanded with $K=100$ principal components from the FPCA in Section \ref{sec:adapt:fpca}. The model is as in \eqref{eq:adapt:funreg}. Since the genetic ancestry was not computed for all individuals, the number of facial manifolds involved in the model is $N=3287$. The model also includes a $\mathcal{N}(0,10)$ noise variable just as a check to make sure our subsequent p-values have proper specificity.

\begin{equation}
\label{eq:adapt:funreg}
\begin{split}
{\bf Y}_n(m) & = {\bf \beta}_0(m) + {\bf \beta}_1(m) \text{sex}_n + {\bf \beta}_2(m) \text{age}_n + {\bf \beta}_3(m) \text{height}_n + {\bf \beta}_4(m) \text{weight}_n\\
&+ {\bf \beta}_5(m) p^\text{E.ASN}_n + {\bf \beta}_6(m) p^\text{S.ASN}_n + {\bf \beta}_7(m) p^\text{AMR}_n + {\bf \beta}_8(m) p^\text{W.AFR}_n + {\bf \beta}_9(m) p^\text{S.EUR}_n\\
& + {\bf \beta}_{10}(m) (\text{sex}_n \times \text{age}_n) + {\bf \beta}_{11}(m) (\text{age}_n \times \text{weight}_n)\\
& + {\bf \beta}_{12}(m) (\text{height}_n \times \text{weight}_n)+ {\bf \beta}_{13}(m) \text{noise}_n + {\bf \epsilon}_n(m).
\end{split}
\end{equation}

We estimated beta functions $\beta_r$'s with regularization term as outlined in Section \ref{sec:method:funreg}. The tuning parameter $\lambda_r$'s are determined using iterative 4-fold cross validation.

The sizes and p-values of resulting $\hat{\beta}_r$ are presented in Table \ref{table:pvals}.  We utilize three different tests as outlined in \cite{choi:reimherr:2016}.  Each test uses slightly different normalizations of the estimated parameter functions.  The first test is based on the $L^2$-norm, which ignores the covariance operator of the parameter estimate (though it is used in calculating p-values).  The other two approaches attempt to normalize by the covariance operator, where the PC and Choi approach normalize by the Moore-Penrose inverse of the covariance operator and square-root of the covariance operator, respectively.  Both approaches can be phrased using PCA, and we refer the interested reader to \cite{choi:reimherr:2016} for more details.

The asymptotic distribution of the PC approach is simply a chi-squared distribution, while the norm and Choi approach are given by weighted sums of chi-squares.  
We approximate p-values from the weighted distribution using Imhof's method \citep{imhof:1961, duchesne:LM:2010}. 
The p-values suggest that all beta functions are significant at 99\% significance level except the noise. Therefore, the tests seem to have discerned the effects from the true negative variable.

\begin{table}[h!]
\begin{center}
\begin{tabular}{| c | c | c | c | c | c |}
\hline
& Predictor & $\| \hat{\beta}_r \|^2$ & p-value (PC) & p-value (Choi) & p-value (Norm)\\ \hline
\hline
$\beta_0$ & & 3.910e-05 & $<$ 1.110e-21 & 5.385e-15 &2.631e-09\\ 
$\beta_1$ & sex & 9.924e-07 & $<$ 1.110e-21 &5.551e-16 &1.443e-15\\ 
$\beta_2$ & age & 4.443e-10 & 5.888e-11& 3.053e-15 &3.672e-04\\ 
$\beta_3$ & height & 2.105e-09 & $<$ 1.110e-21 &5.551e-17 &3.514e-14 \\ 
$\beta_4$ & weight & 1.549e-09&  9.826e-08 &6.088e-06 &1.173e-02 \\ 
$\beta_5$ & $p^\text{E.ASN}$ & 2.439e-06 & $<$ 1.110e-21 &4.996e-15 &3.164e-15 \\ 
$\beta_6$ & $p^\text{S.ASN}$ & 5.859e-07 & 1.332e-17 &2.220e-16 &1.720e-11 \\ 
$\beta_7$ & $p^\text{AMR}$ & 7.689e-07 & 1.418e-21 &1.110e-16 &1.357e-08 \\ 
$\beta_8$ & $p^\text{W.AFR}$ & 1.783e-06 & $<$ 1.110e-21 &1.110e-15 &7.772e-16 \\ 
$\beta_9$ & $p^\text{S.EUR}$ & 1.514e-06 & $<$ 1.110e-21 &6.661e-16 &8.826e-14 \\ 
$\beta_{10}$ & sex $\times$ age & 2.103e-10 & 1.988e-19 &5.551e-17 &2.034e-10 \\ 
$\beta_{11}$ & age $\times$ weight & 3.523e-14&  7.790e-06& 5.328e-11 &9.143e-03 \\ 
$\beta_{12}$ & height $\times$ weight & 7.553e-14 & 3.521e-10 &5.074e-07 &1.488e-03\\ 
$\beta_{13}$ & noise  & 4.188e-12  &3.419e-01& 3.242e-01& 5.647e-01\\ 
\hline
\end{tabular}
\label{table:pvals}
\caption{The size of $\hat{\beta}_r$ and p-values based on PC test, Choi test, and Norm test are presented.}
\end{center}
\end{table}

Now that we have carried out our hypothesis testing, it is important to visualize and more fully understand the estimated beta functions.  Since these functions have domain of ${\bf M}_0$, plotting $\beta_r$ is challenging. Instead, we visualize the effects in a manner similar to the PC functions in Section \ref{sec:adapt:fpca}; at each point we examine how strong the effect is in the orthogonal direction to the tangent plane (i.e. outward or inward relative to the face).

We provide three different plot types: directional plots, pointwise significance plots, and overall significance plots that control the Type 1 error rate simultaneously across the face. The directional plot shows how the beta function affects the face, the pointwise significance plot shows the facial areas where each point is tested positive (blue/red means positive at 99\% level and lightblue/orange means positive at 95\% level), and the overall significance plot shows the facial areas that have overall significance at 99\% level. An advantage of applying functional data analysis tools to faces are these overall significance plots, which rely heavily on the functional nature of the data. 

We discuss only a few of the estimated effects here to highlight how to interpret our results.  Further discussion can be found in the supplemental.  For example, the middle plot of Figure \ref{fig:sex_effect} presents the effect for sex, demonstrating the difference between the average male and female face, for a subject that is 30 years old, 170cm tall, and weighs 70kg. Blue denotes an inward effect, while red represents an outward effect. The female face is rounder than the male's as signified by the red parts around the cheek in the directional plot. It also shows that the male has a more pronounced nose, and the female has a rounder eye area with red on the eyelids and blue on the eyebrow area. Those areas are shown as significant for both pointwise significance plot and overall significance plot.

\begin{figure}[ht]
\begin{center}
\includegraphics[width = .95\linewidth]{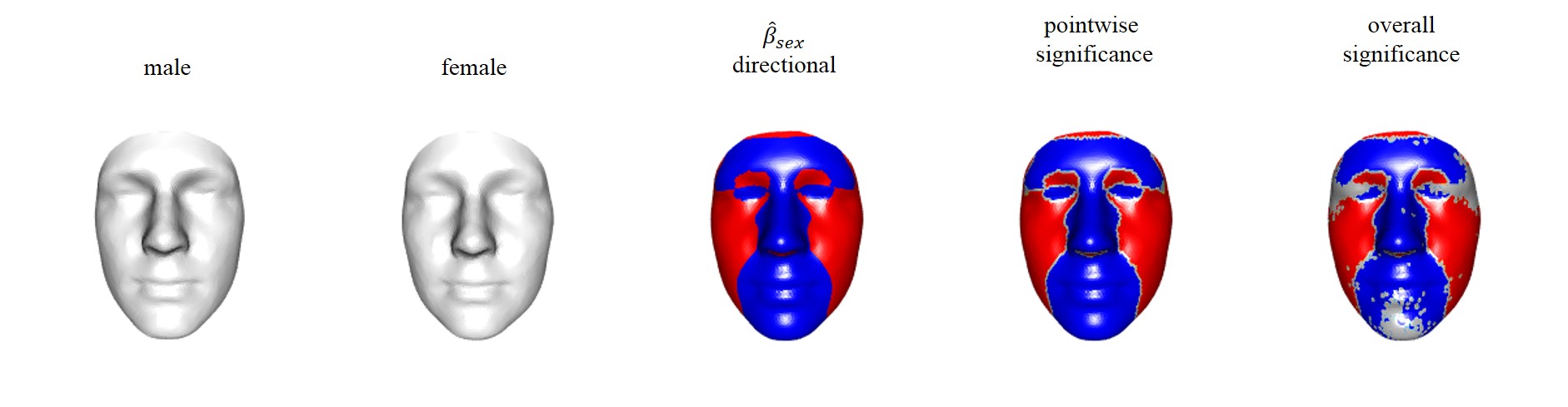}
\caption{The left two plots are predicted faces of 30-year-old, 170cm-tall, 70kg-heavy Northern European male and female. The right three plots show the effect of beta of sex.}
\label{fig:sex_effect}
\end{center}
\end{figure}

The effect of the proportion of East Asian descent is shown in Figure \ref{fig:ASN_effect}. As Northern European proportion is taken as the base, the beta function indicates the difference between Northern European and East Asian. We see that the average East Asian face is rounder, has a lower nose, and a less pronounced eyebrow.  The overall significance plot (right most plot) shows that the nose, cheek, and forehead area are still statistically significant at a 99\% significance level when correcting for multiple testing across the entire face using our confidence bubbles.

\begin{figure}[ht]
\begin{center}
\includegraphics[width = .95\linewidth]{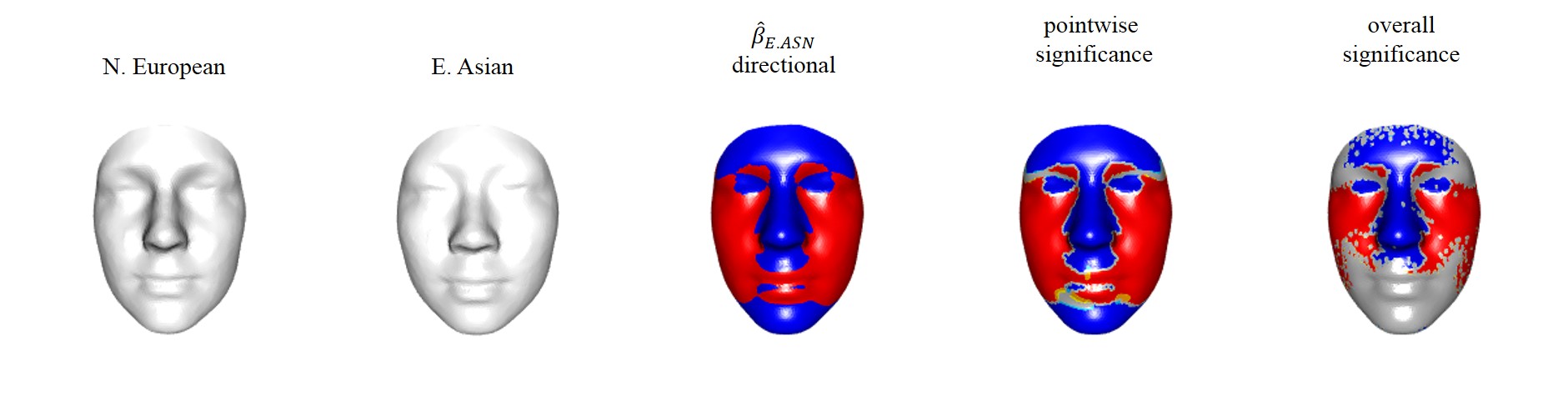}
\caption{The left two plots are predicted faces of 25-year-old, 165cm-tall, 65kg-heavy Northern European and East Asian female. The right three plots show the effect of the corresponding beta.}
\label{fig:ASN_effect}
\end{center}
\end{figure}

The effect of the proportion of Western African is shown in Figure \ref{fig:AFR_effect}. The most features seem to be the nose and mouth, and those are picked up in the overall significance plot. The nose of Western African is more flattened but more wider than that of Northern European, shown as the inward effect (blue) in the middle of nose, and the outward effect (red) on the sides of nose. The lips are more outward, and the lower cheek area difference is also picked up in the overall significance plot.

\begin{figure}[ht]
\begin{center}
\includegraphics[width = .95\linewidth]{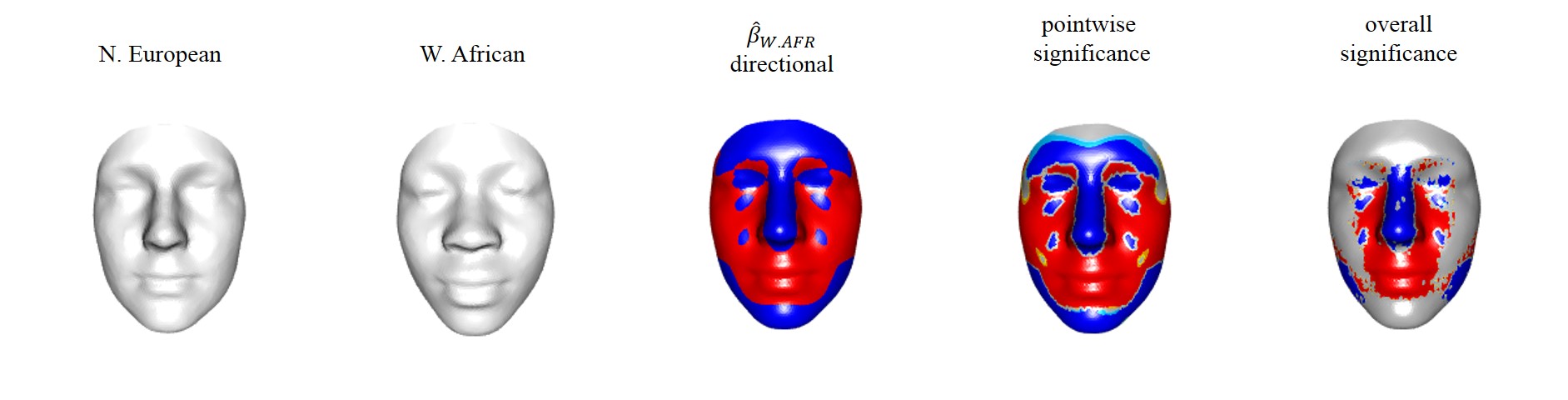}
\caption{The left two plots are predicted faces of 25-year-old, 165cm-tall, 65kg-heavy Northern European and Western African female. The right three plots show the effect of beta of the corresponding beta.}
\label{fig:AFR_effect}
\end{center}
\end{figure}

\section{Discussion}
\label{sec:discussion}

We have presented a new framework that facilitates the analysis of 3D imaging data sampled at a high-resolution. We view Manifold Data Analysis (MDA) as a subbranch of Functional Data Analysis (FDA), which  incorporates ideas and techniques from shape analysis and manifold learning. MDA allows a wide variety of functional data analysis tools to be applied to data that includes a manifold variable by converting each manifold unit into a functional object. Through this method, the dimension of the data is reduced and smoothness of data can be exploited. Extensions of functional data techniques are introduced to the manifold setting: 2-step functional principal component analysis and manifold-on-scalar regression. Our methods are illustrated using 3D facial images.

This paper opens up a broad range of future work. In constructing functional objects, there are many choices to be made including the manifold learning technique for domain construction, the basis system for functional unit construction, and the tuning parameter selection. Here we used the FELSPLINE basis which also leads to further issues of how to construct the domain mesh. 
Alternative basis functions could prove useful, especially kernel methods that allow for irregular multidimensional domains as well as multivariate functions.  
Also, other extensions of functional data techniques for manifold data can be developed. 

Lastly, we believe that these techniques will prove useful to a variety of applications.  As part of the big data revolution occurring in the sciences, many types of high-frequency or high-resolution data are being collected.  Data which include samples of manifolds will become increasingly common, especially in biomedical imaging.  FDA tools naturally exploit smoothness and we thus believe the will be useful for analyzing data involving manifolds.  
As computational tools progress, we will be likely able to work with deformation maps that directly map one manifold to another, even if those manifolds reside in 3D space.  However, the computational challenges may be substantial, especially if more complicated statistical models are to be employed. For example, building nonlinear or varying coefficient models where the outcome is a manifold. Or, as we aim to do in a future work, if one wants to carry out a regression with a large number predictors (e.g. genetic markers).  There are many open practical and methodological issues that remain to be explored. 

\bibliographystyle{abbrvnat}
{\renewcommand{\baselinestretch}{0.9}
\small
\bibliography{biblio}
}

\clearpage

\appendix
\setcounter{page}{1}
\renewcommand\thefigure{\thesection.\arabic{figure}}  
\renewcommand\thetable{\thesection.\arabic{table}}
\setcounter{figure}{0}  
\begin{center}
\bf{\large Online Supplemental Material}
\end{center}

\section{2-Step Functional Principal Component Analysis}
\paragraph*{Step 1.} 
Without loss of generality, assume that the ${\bf Y}_n(m)$ have been centered and thus have mean zero.   Each of the functions is expressed as
$$
{\bf Y}_{n}(m) = \left[ Y_{n1}(m), \cdots, Y_{nD}(m) \right]^\top \approx\left[
\sum_{j = 1}^J \hat b_{nj1} e_j(m), \cdots, \sum_{j = 1}^J \hat b_{njD} e_j(m)
\right]^\top,
$$
for $n = 1, \dots, N$. We stack all of the coordinate-wise functions into a single vector of functions with dimension $N D$.  We denote the resulting functions as $Y_l(m)$  where $Y_l(m) = Y_{n q}(m)$ for $l = N(q-1)+n$ and $n = 1, \cdots, N$, $q = 1, \cdots, D$.

We aim to find the pairs of eigenvalues $\eta_h$ and principal component functions $\psi_h: \bM_0 \to \mbR^D$, for $h=1,\dots, H$, which satisfy
$$ \eta_h \psi_h(m) = \int \Phi(m, m') \psi_h(m') dm' $$ 
where 
\begin{align*}
\bPhi(m,m') = E[Y_l(m)Y_l(m')^\top] &= \sum_{j_1}^J \sum_{j_2}^J E[b_{lj_1} b_{lj_2}] e_{j_1}(m) e_{j_2}(m')^\top\\
&= \sum_{j_1}^J \sum_{j_2}^J \Pi_{j_1, j_2} e_{j_1}(m) e_{j_2}(m')^\top,
\end{align*}
and $\|\psi_h\|=1$.\\
We expand $\psi_h$ using $e_j$:
$$\psi_h(m) = \sum_{j=1}^J w_{hj} e_j(m).$$
We then need to solve the following system of linear equations
\begin{align*}
\eta_h \sum_{j=1}^J w_{hj} e_j(m) &= \int_{{\bf M}_0} \left( \sum_{j_1}^J \sum_{j_2}^J \Pi_{j_1, j_2} e_{j_1}(m) e_{j_2}(m') \right) \left( \sum_{j_3}^J w_{hj_3} e_{j_3}(m') \right) dm'\\
&=\sum_{j_1}^J \sum_{j_2}^J \sum_{j_3}^J \Pi_{j_1, j_2} \left( \int_{{\bf M}_0} e_{j_2}(m') e_{j_3}(m') dm'\right) e_{j_1}(m)\\
&=\sum_{j_1}^J \sum_{j_2}^J \sum_{j_3}^J \Pi_{j_1, j_2} \bZ_{j_2, j_3} e_{j_1}(m).
\end{align*}
And $\|\psi_h\|=1$ means
$$ \int_{{\bf M}_0} \sum_{j_1 = 1}^J \sum_{j_2 = 1}^J w_{hj_1} e_{j_1}(m) w_{hj_2} e_{j_2}(m) dm= \sum_{j_1 = 1}^J \sum_{j_2 = 1}^J w_{hj_1} w_{hj_2} \bZ_{j_1, j_2} = 1.$$
Factor the matriz $\bZ = \bG^\top \bG$ so that
$$\sum_{j_1 = 1}^J \sum_{j_2 = 1}^J w_{hj_1} w_{hj_2} \bZ_{j_1, j_2} = \sum_{j_1=1}^J \sum_{j_2=1}^J \sum_{j_3=1}^J w_{hj_1} w_{hj_2} {\bf G}_{j_1 j_3} {\bf G}_{j_3 j_2} = \sum_{j=1}^J a_{hj}^2,$$
where we define $a_{hj} = \sum_{j_1=1}^J w_{h j_1} {\bf G}_{j j_1}$.   We then have
\begin{align*}
\sum_{j_1}^J \sum_{j_2}^J \sum_{j_3}^J \Pi_{j_1, j_2} \bZ_{j_2, j_3} e_{j_1}(m) 
&= \sum_{j_1}^J \sum_{j_2}^J \sum_{j_3}^J \sum_{j_4}^J \Pi_{j_1 j_2} {\bf G}_{j_2 j_4} {\bf G}_{j_4 j_3} w_{h j_3} e_{j_1} (m)\\
&= \sum_{j_1}^J \sum_{j_2}^J \sum_{j_4}^J \Pi_{j_1 j_2} {\bf G}_{j_2 j_4} a_{h j_4} e_{j_1} (m)\\
&= \sum_{j_1}^J \sum_{j_4}^J \left( \sum_{j_2}^J \Pi_{j_1 j_2} {\bf G}_{j_2 j_4} \right) a_{h j_4} e_{j_1} (m).
\end{align*} 
So we obtain the relation
$$\eta_h w_{h j_1} = \sum_{j_2}^J \left( \sum_{j_3}^J \Pi_{j_1 j_3} {\bf G}_{j_3 j_2} \right) a_{h j_2}$$ and
$$\eta_h a_{h j} = \eta_h \sum_{j_1}^J w_{h j_1} {\bf G}_{j, j_1} = \sum_{j_1}^J \sum_{j_2}^J {\bf G}_{j, j_1} \left( \sum_{j_3}^J \Pi_{j_1 j_3} {\bf G}_{j_3 j_2} \right) a_{h j_2} = \sum_{j_2}^J \tilde{\Pi}_{j, j_2} a_{h j_2},$$
where $\tilde{\Pi}_{j, j_2} = \sum_{j_1}^J \sum_{j_3}^J {\bf G}_{j, j_1} \Pi_{j_1 j_3} {\bf G}_{j_3 j_2}$.  
So we the vector ${\bf a}_j = \{a_{h j}\}$ is the $j^{th}$ eigenvector of the covariance matrix $\tilde{\Pi}$. 
Since we know $$a_{hj} = \sum_{j_1=1}^J w_{h j_1} {\bf G}_{j j_1},$$
reversing it would give $w_{h j_1}$ and then we can get $$\psi_h(m) = \sum_{j=1}^J w_{hj} e_j(m).$$\\

\paragraph*{Step 2.} We now expand $Y_{nq}(m)$ using the $\{\psi_h(m)\}$: 
$$Y_{nq}(m) = \sum_{h=1}^H c_{nhq} \psi_h(m).$$
The coefficients ${\bf c} = \{c_{nhq}\}$ form an $N \times H \times D$ array. The covariance operator of ${\bf Y}_n(m)$ is given by 
\begin{align*}
\Gamma_{q, q'} (m, m') = E[Y_{nq}(m) Y_{nq'}(m')] &= \sum_{h_1=1}^H \sum_{h_2=1}^H E[c_{n h_1 q} c_{n h_2 q'}^\top] \psi_{h_1}(m) \psi_{h_2}(m')\\
&= \sum_{h_1=1}^H \sum_{h_2=1}^H \Sigma_{h_1 q h_2 q'} \psi_{h_1}(m) \psi_{h_2}(m').
\end{align*}
Now we find the pairs of eigenvalues $\lambda_k$ and eigenfunctions $\bV_k(m)$ that satisfy
$$\lambda_k \bV_k(m) = \int \bGa(m,m') \bV_k(m') dm'$$ 
where $\|\bV_k\|=1$. We expand $\bV_k$ using the $\psi_h$ as well:
$$V_{kq}(m) = \sum_{h=1}^H v_{khq} \psi_h(m),$$
where ${\bf v} = \{v_{khq}\}$ is a $K \times H \times D$ array of coefficients. So we want to solve
\begin{align*}
\lambda_k \sum_{h=1}^H v_{khq} \psi_h(m) 
&= \sum_{q'=1}^D \int_{{\bf M}_0} \sum_{h_1=1}^H \sum_{h_2=1}^H \Sigma_{h_1 q h_2 q'} \psi_{h_1}(m) \psi_{h_2}(m') \sum_{h_3=1}^H v_{k h_3 q'} \psi_{h_3}(m') dm'\\
&=\sum_{q'=1}^D \sum_{h_1=1}^H \sum_{h_2=1}^H \sum_{h_3=1}^H \Sigma_{h_1 q h_2 q'} \left( \int_{{\bf M}_0} \psi_{h_2}(m') \psi_{h_3}(m') dm' \right) v_{k h_3 q'} \psi_{h_1}(m)\\
&=\sum_{q'=1}^D \sum_{h_1=1}^H \sum_{h_2=1}^H \sum_{h_3=1}^H \Sigma_{h_1 q h_2 q'} {\bf W}_{h_2 h_3} v_{k h_3 q'} \psi_{h_1}(m)\\
&=\sum_{q'=1}^D \sum_{h_1=1}^H \sum_{h_2=1}^H \Sigma_{h_1 q h_2 q'} v_{k h_2 q'} \psi_{h_1}(m),
\end{align*}
since $\bW$ is the identity matrix as the $\psi_h$ are orthogonal. Since $\|\bV_k|=1$ this means that
$$\sum_{q=1}^D \sum_{h_1=1}^H \sum_{h_2=1}^H v_{k h_1 q} v_{k h_2 q} = 1.$$
Therefore
$$\lambda_k v_{k h q} = \sum_{q'=1}^D \sum_{h_2 = 1}^H \Sigma_{h q h_2 q'} v_{k h_2 q'}.$$
So we have that ${\bf v}_k = \{v_{k h q}\}$ is the $k^{th}$ eigenmatrix of the $H \times D \times H \times D$ covariance tensor $\Sigma$.

\section{Manifold-on-Scalar Regression with Regularization}
Our objective is to find $\beta$'s that minimizes 
\begin{equation}
\label{eq:obj}
\sum_{n=1}^N \int_{{\bf M}_0} \left| {\bf Y}_n (m) - \sum_{r=1}^R x_{nr} {\bf \beta}_r (m) \right|^2 dm + \sum_{r=1}^R \lambda_r \int_{{\bf M}_0} |L {\bf \beta}_r (m)|^2 dm
\end{equation}

We can take $\lambda = \lambda_1 = \cdots = \lambda_R$, but we will keep them separate for now.\\

We need to choose roughness operator $L$ carefully. Since we do not want the minimizer of \ref{eq:obj} to change depending on the coordinate system, we need an operator that is invariant to rotation and translation. Ramsey (2002) chooses Laplacian operator as such operator. Laplacian operator $\bigtriangleup$ is such that $\bigtriangleup f = f_{xx} + f_{yy}$.

Wtih ${\bf f}: [{\bf M}_0 \subset \mathbb{R}^2] \rightarrow \mathbb{R}^3$,  the Laplacian operator on ${\bf f}$ is as $$\bigtriangleup {\bf f} = \bigtriangleup [f_1, f_2, f_3]^\top = [\bigtriangleup f_1, \bigtriangleup f_2, \bigtriangleup f_3]^\top = \left[\frac{d^2 f_1}{d m_1^2} + \frac{d^2 f_1}{d m_2^2}, \frac{d^2 f_2}{d m_1^2} + \frac{d^2 f_2}{d m_2^2}, \frac{d^2 f_3}{d m_1^2} + \frac{d^2 f_3}{d m_2^2} \right]^\top$$ where $f_1, f_2, f_3$ correspond to each coordinate of ${\bf f}$ and $m_1$ and $m_2$ correspond to each coordinate of ${\bf M}_0$.

Therefore \ref{eq:obj} becomes
\begin{equation}
\label{eq:obj2}
\sum_{n=1}^N \int_{{\bf M}_0} \left| {\bf Y}_n (m) - \sum_{r=1}^R x_{nr} {\bf \beta}_r (m) \right|^2 dm + \sum_{r=1}^R \lambda_r \int_{{\bf M}_0} |\bigtriangleup {\bf \beta}_r (m)|^2 dm
\end{equation}

With PC basis functions ${\bf V}_k: [{\bf M}_0 \subset \mathbb{R}^2] \rightarrow \mathbb{R}^3$ for $k = 1, \cdots, K$, ${\bf Y}_n$ and ${\bf \beta}_r$ can both be expanded with ${\bf V}_1, \cdots, {\bf V}_K$:
\begin{equation}
\label{eq:obj3}
\sum_{n=1}^N \int_{{\bf M}_0} \left| \sum_{k=1}^K y_{nk} {\bf V}_k (m) - \sum_{r=1}^R x_{nr} \sum_{k=1}^K b_{rk} {\bf V}_k (m) \right|^2 dm + \sum_{r=1}^R \lambda_r \int_{{\bf M}_0} \left| \sum_{k=1}^K b_{rk}  ( \bigtriangleup {\bf V}_k (m) ) \right|^2 dm
\end{equation}

The first term
\begin{align*}
&\sum_{n=1}^N \int_{{\bf M}_0} \left| \sum_{k=1}^K y_{nk} {\bf V}_k - \sum_{r=1}^R x_{nr} \sum_{k=1}^K b_{rk} {\bf V}_k \right|^2 dm\\
&=\sum_{n=1}^N \int_{{\bf M}_0} \left(\sum_{k=1}^K y_{nk} {\bf V}_k - \sum_{r=1}^R x_{nr} \sum_{k=1}^K b_{rk} {\bf V}_k \right)^\top \left(\sum_{k=1}^K y_{nk} {\bf V}_k - \sum_{r=1}^R x_{nr} \sum_{k=1}^K b_{rk} {\bf V}_k  \right) dm\\
&=\sum_{n=1}^N \int_{{\bf M}_0} \left(\sum_{k=1}^K y_{nk} {\bf V}_k \right)^\top \left(\sum_{k=1}^K y_{nk} {\bf V}_k \right) - \left( \sum_{r=1}^R x_{nr} \sum_{k=1}^K b_{rk} {\bf V}_k  \right)^\top \left(\sum_{k=1}^K y_{nk} {\bf V}_k \right) \\
&\quad - \left(\sum_{k=1}^K y_{nk} {\bf V}_k \right)^\top \left( \sum_{r=1}^R x_{nr} \sum_{k=1}^K b_{rk} {\bf V}_k  \right) + \left( \sum_{r=1}^R x_{nr} \sum_{k=1}^K b_{rk} {\bf V}_k  \right)^\top \left( \sum_{r=1}^R x_{nr} \sum_{k=1}^K b_{rk} {\bf V}_k  \right) dm\\
&=\sum_{n=1}^N \sum_{k_1=1}^K \sum_{k_2=1}^K y_{nk_1} y_{nk_2} \int_{{\bf M}_0} {\bf V}_{k_1}^\top {\bf V}_{k_2} dm - 2 \sum_{n=1}^N \sum_{r=1}^R x_{nr} \sum_{k_1=1}^K \sum_{k_2=1}^K b_{rk_1} y_{nk_2} \int_{{\bf M}_0} {\bf V}_{k_1}^\top {\bf V}_{k_2} dm \\
&\quad + \sum_{n=1}^N \sum_{r_1=1}^R\sum_{r_2=1}^R x_{nr_1} x_{nr_2} \sum_{k_1=1}^K \sum_{k_2=1}^K b_{r_1 k_1} b_{r_2 k_2}\int_{{\bf M}_0} {\bf V}_{k_1}^\top {\bf V}_{k_2} dm\\
&=\sum_{n=1}^N \sum_{k=1}^K y_{nk}^2 - 2 \sum_{n=1}^N \sum_{r=1}^R x_{nr} \sum_{k=1}^K  b_{rk} y_{nk} + \sum_{n=1}^N \sum_{r_1=1}^R\sum_{r_2=1}^R x_{nr_1} x_{nr_2} \sum_{k=1}^K b_{r_1 k} b_{r_2 k}\\
&=\sum_{n=1}^N \sum_{k=1}^K \left( y_{nk}^2 - 2 \sum_{r=1}^R x_{nr} b_{rk} y_{nk} + \sum_{r_1=1}^R\sum_{r_2=1}^R x_{nr_1} x_{nr_2}  b_{r_1 k} b_{r_2 k} \right)^2\\
&=\sum_{n=1}^N \sum_{k=1}^K \left( y_{nk} - \sum_{r=1}^R x_{nr} b_{rk} \right)^2
\end{align*}

The second term
\begin{align*}
&\sum_{r=1}^R \lambda_r \int_{{\bf M}_0} \left| \sum_{k=1}^K b_{rk}  ( \bigtriangleup {\bf V}_k ) \right|^2 dm\\
&=\sum_{r=1}^R \lambda_r \int_{{\bf M}_0} \left( \sum_{k=1}^K b_{rk}  ( \bigtriangleup {\bf V}_k ) \right)^\top \left( \sum_{k=1}^K b_{rk}  ( \bigtriangleup {\bf V}_k ) \right) dm\\
&=\sum_{r=1}^R \lambda_r \sum_{k_1=1}^K \sum_{k_2=1}^K b_{rk_1} b_{rk_2} \int_{{\bf M}_0} ( \bigtriangleup {\bf V}_{k_1} )^\top ( \bigtriangleup {\bf V}_{k_2} ) dm\\
&=\sum_{r=1}^R \lambda_r \sum_{k_1=1}^K \sum_{k_2=1}^K b_{rk_1} b_{rk_2} U_{{k_1}, {k_2}}
\end{align*}
where 
\begin{align*} 
U_{{k_1}, {k_2}} &= \int_{{\bf M}_0} ( \bigtriangleup {\bf V}_{k_1} )^\top ( \bigtriangleup {\bf V}_{k_2} ) dm\\
&= \int_{{\bf M}_0} (\bigtriangleup V_{k_1, 1})^2 + (\bigtriangleup V_{k_1, 2})^2 + (\bigtriangleup V_{k_1, 3})^2 dm.
\end{align*}
that is summing the three coordinates.\\
And from FPCA, we know $$V_{kq}(m) = \sum_{h=1}^H v_{khq} \phi_h(m)$$ and $$\phi_h(m) = \sum_{j=1}^J w_{hj} e_j(m)$$
where $e_j$'s are the FELSPLINEs (Ramsey, 2002).\\
Then $$\int_{{\bf M}_0} (\bigtriangleup V_{k, q} (m) )^2 dm = \int_{{\bf M}_0} \left(\sum_{h=1}^H (\bigtriangleup \phi_h(m) \right)^2 dm.$$
And in order to get $\int_{{\bf M}_0} (\bigtriangleup \phi_h(m))^2 dm$, we need to consider the FEM theories.

Let $f_h(m) = - \bigtriangleup \phi_h(m)$.
\begin{align*}
\langle f_h, e_j \rangle &= \int_{{\bf M}_0} (- \bigtriangleup \phi_h) e_j dm\\
&= \int_{{\bf M}_0} (\bigtriangledown \phi_h) (\bigtriangledown e_j ) \quad (\because \text{ Green's theorem})\\
&= \int_{{\bf M}_0} (\bigtriangledown \sum_{j_1=1}^J w_{hj_1} e_{j_1}) (\bigtriangledown e_j )\\
&= \sum_{j_1=1}^J w_{hj_1} \int_{{\bf M}_0} (\bigtriangledown e_{j_1})(\bigtriangledown e_j) dm
\end{align*}
And we have code for getting $\int_{{\bf M}_0} (\bigtriangledown e_{j_1})(\bigtriangledown e_j) dm$. The matrix with these components is called stiffness matrix. \\
Then using
$$f_h(m) = \sum_{j=1}^J \langle f_h, e_j \rangle e_j(m) = \sum_{j=1}^J f_{hj} e_j(m),$$ we can get
\begin{align*}
\int_{{\bf M}_0} (\bigtriangleup \phi_h)^2 dm &= \int_{{\bf M}_0} f_h(m)^2 dm \\
&= \int_{{\bf M}_0} (\sum_{j_1=1}^J f_{h j_1} e_{j_1}(m)) (\sum_{j_2=1}^J f_{h j_2} e_{j_2}(m)) dm\\
&= \sum_{j_1=1}^J \sum_{j_2=1}^J  f_{h j_1} f_{h j_2} \int_{{\bf M}_0} e_{j_1}(m)  e_{j_2}(m) dm.
\end{align*}
The matrix with components of the inner product between $e_{j_1}$ and $e_{j_2}$ ($\int_{{\bf M}_0} e_{j_1}(m)  e_{j_2}(m) dm$) is called mass matrix, and we have code for that too.\\\\

Let's go back to getting the least square estimate of $\{ b_{rk} \}$.\\
Let $$Y = \left[ \begin{matrix} y_{11} & y_{12} & \cdots & y_{1K}\\ y_{21} & y_{22} & \cdots & y_{2K} \\ \vdots & \vdots & \vdots & \vdots\\ y_{N1} & y_{N2} & \cdots & y_{NK} \end{matrix} \right], \quad 
X = \left[ \begin{matrix} x_{11} & x_{12} & \cdots & x_{1R}\\ x_{21} & x_{22} & \cdots & x_{2R} \\ \vdots & \vdots & \vdots & \vdots\\ x_{N1} & x_{N2} & \cdots & x_{NR} \end{matrix} \right],$$
$$B = \left[ \begin{matrix} b_{11} & b_{12} & \cdots & b_{1K}\\ b_{21} & b_{22} & \cdots & b_{2K} \\ \vdots & \vdots & \vdots & \vdots\\ b_{R1} & b_{R2} & \cdots & b_{RK} \end{matrix} \right], \quad 
\Lambda = diag(\lambda_1, \lambda_2, \cdots, \lambda_R).$$

Then our objective becomes to find $B$ that minimizes
\begin{equation}
\label{eq:obj4}
trace\{(Y - XB)^\top (Y - XB)\} + trace\{\Lambda B U B^\top \}
\end{equation}

\bigskip \bigskip

Now let's find the least square estimate of $B$.\\
Differentiate \ref{eq:obj4} and set it to 0: $${-2X^\top Y} + 2 X^\top X B + 2 \Lambda B U = 0.$$
We can cross 2 out: $${- X^\top Y} + X^\top X B + \Lambda B U = 0.$$
Take transpose of everything: $${- Y^\top X} + B^\top (X^\top X) + U^\top B^\top \Lambda = 0.$$
Vectorize the whole thing: $${- \text{vec}(Y^\top X)} + ((X^\top X) \otimes I_K) \text{vec}(B^\top) + (\Lambda \otimes U^\top) \text{vec}(B^\top) = 0.$$
Then the least square estimate of $B$ is:
\begin{equation}
\label{eq:lse}
\text{vec}(\hat{B}^\top) = \left( (X^\top X) \otimes I_K + \Lambda \otimes U^\top \right)^{-1} \text{vec}(Y^\top X).
\end{equation}

Find covariance of $\text{vec}(\hat{B}^\top)$. Let $$A = \left( (X^\top X) \otimes I_K + \Lambda \otimes U^\top \right)^{-1}.$$
\begin{align*}
\text{cov} \left( \text{vec}(\hat{B}^\top) \right) &= A \text{cov} \left( \text{vec}(Y^\top X) \right) A^\top\\
&=A \text{cov} \left( (X^\top \otimes I_K) \text{vec}(Y^\top) \right) A^\top\\
&= A (X^\top \otimes I_K) (I_N \otimes \Sigma) (X \otimes I_K) A^\top\\
&= A (X^\top \otimes \Sigma) (X \otimes I_K) A^\top\\
&= A \left( (X^\top X) \otimes \Sigma \right) A^\top
\end{align*}

\section{Proof of Theorem \ref{t:bubble}}
Using the Karhunen-Lo\`{e}ve (KL) expansion, we can write $$\sqrt{N}(\hat{\beta}_r(m) - \beta_r(m)) = \sum_{j=1}^\infty \sqrt{\lambda_j} Z_j {\bf U}_j(m),$$
where the equality hold for almost all $m \in \bM_0$ since $L^2(\bM_0)$ consists of equivalence classes.
By the Cauchy-Schwarz inequality,
\begin{align*}
N|\hat{\beta}_r(m) - \beta_r(m) |^2 &\leq \sum_{j=1}^\infty |\lambda_j^{\frac{1}{4}} Z_j |^2 \sum_{j=1}^\infty | \lambda_j^{\frac{1}{4}} {\bf U}_j(m) |^2\\
&= \sum_{j=1}^\infty \sqrt{\lambda_j} Z^2_j \sum_{j=1}^\infty \sqrt{\lambda_j}|{\bf U}_j(m) |^2,
\end{align*}
as desired.

\end{document}